\documentclass[%
 reprint,
superscriptaddress,
showpacs,showkeys,preprintnumbers,
 amsmath,amssymb,
 aps,
]{revtex4-1}

\usepackage{graphicx}
\usepackage{dcolumn}
\usepackage{bm}
\usepackage{here}
\usepackage[dvips]{color}
\usepackage{multirow}
\usepackage[columnwise]{lineno}

\newcommand{\diff}[2]{ \frac{{\mathrm{d}}{#1}}{{\mathrm{d}}{#2}} }

\newcommand{\pdiff}[2]{ \frac{\partial{#1}}{\partial{#2}} }

\newcommand{\abs}[1]{\left\vert{#1}\right\vert}

\newcommand{\sixJ}[6]{ \left\{ \! \begin{array}{ccc} {#1}&{\! #2}&{\! #3}\\{#4}&{\! #5}&{\! #6} \end{array} \! \right\} }
\newcommand{\nineJ}[9]{ \left\{ \! \begin{array}{ccc} {#1}&{\! #2}&{\! #3}\\{#4}&{\! #5}&{\! #6}\\{#7}&{\! #8}&{\! #9} \end{array} \! \right\} }

\newcommand{\asym}[2]{\frac{{#1}-{#2}}{{#1}+{#2}}}

\makeatletter
  \def\erf{\mathop{\operator@font erf}\nolimits}
  \def\erfc{\mathop{\operator@font erfc}\nolimits}
  \def\Erf{\mathop{\operator@font Erf}\nolimits}
  \def\Shi{\mathop{\operator@font Shi}\nolimits}
  \def\Chi{\mathop{\operator@font Chi}\nolimits}
  \def\Ei{\mathop{\operator@font Ei}\nolimits}
  \def\cosec{\mathop{\operator@font cosec}\nolimits}
  \def\sech{\mathop{\operator@font sech}\nolimits}
  \def\cosech{\mathop{\operator@font cosech}\nolimits}
  \newcommand\hypgeo[2]{{}_{#1}{\operator@font F}_{#2}}
  \def\Re{\mathop{\operator@font Re}\nolimits}
  \def\Im{\mathop{\operator@font Im}\nolimits}

\makeatother

\begin{document}


\title{Angular Distribution of $\gamma$-rays from Neutron-Induced Compound States of $^{140}$La}

\def\affNagoya{Nagoya University, Furocho, Chikusa, Nagoya 464-8602, Japan}
\def\affKyushu{Kyushu University, 744 Motooka, Nishi, Fukuoka 819-0395, Japan}
\def\affJAEA{Japan Atomic Energy Agency, 2-1 Shirane, Tokai 319-1195, Japan}
\def\affDenso{Denso Corporation, 1-1, Showa-cho, Kariya, Aichi 448-8661, Japan}
\def\affMusen{New Japan Radio Corporation, 3-10, Nihonbashi Yokoyama-cho,Chuo-ku, Tokyo 103-8456, Japan}

\author{T.~Okudaira}
\affiliation{\affNagoya}

\author{S.~Takada}
\affiliation{\affKyushu}


\author{K.~Hirota}
\affiliation{\affNagoya}

\author{A.~Kimura}
\affiliation{\affJAEA}

\author{M.~Kitaguchi}
\affiliation{\affNagoya}

\author{J.~Koga}
\affiliation{\affKyushu}

\author{K.~Nagamoto}
\thanks{Present Address: \affDenso}
\affiliation{\affNagoya}

\author{T.~Nakao}
\thanks{Present Address: \affNagoya}
\affiliation{\affJAEA}

\author{A.~Okada}
\thanks{Present Address: \affMusen}
\affiliation{\affNagoya}

\author{K.~Sakai}
\affiliation{\affJAEA}

\author{H.~M.~Shimizu}
\affiliation{\affNagoya}

\author{T.~Yamamoto}
\affiliation{\affNagoya}

\author{T.~Yoshioka}
\affiliation{\affKyushu}


\date{\today}

\begin{abstract}
Angular distribution of individual $\gamma$-rays, emitted from a neutron-induced compound nuclear state via radiative capture reaction of ${}^{139}$La(n,$\gamma$) has been studied as a function of incident neutron energy in the epithermal region by using germanium detectors.

An asymmetry $A_{\mathrm{LH}}$ was defined as $(N_{\mathrm L}-N_{\mathrm H})/(N_{\mathrm L}+N_{\mathrm H})$, where $N_{\mathrm L}$ and $N_{\mathrm H}$ are integrals of low and high energy region of a neutron resonance respectively, and we found that $A_{\mathrm{LH}}$ has the angular dependence of $(A\cos\theta_\gamma+B)$, where $\theta_\gamma$ is emitted angle of $\gamma$-rays, with $A= -0.3881\pm0.0236$ and  $B=-0.0747\pm0.0105$ in 0.74~eV p-wave resonance. 

This angular distribution was analyzed within the framework of interference between s- and p-wave amplitudes in the entrance channel to the compound nuclear state, and it is interpreted as the value of the partial p-wave neutron width corresponding to the total angular momentum of the incident neutron combined with the weak matrix element, in the context of the mechanism of enhanced parity-violating effects.
Additionally we used the result to quantify the possible enhancement of the breaking of the time-reversal invariance in the vicinity of the p-wave resonance.
\end{abstract}

\pacs{13.75.Cs, 
21.10.Hw,
21.10.Jx,
21.10.Re,
23.20.En,
24.30.Gd,
24.80.+y,
25.40.Fq,
25.70.Gh,
27.60.+j,
29.30.Kv
}
\keywords{compound nuclei,
partial wave interference,
neutron radiative capture reaction,
discrete symmetry}
\maketitle


\section{Introduction}
The magnitude of parity violating effects in effective nucleon--nucleon interactions is $10^{-7}$, as observed in the helicity dependence of the total cross section between nucleons~\cite{pot74,yua86,ade85}.
Extremely large parity violation(P-violation) was found in the helicity dependence of the neutron absorption cross section in the vicinity of p-wave resonance of $^{139}$La+n~\cite{alf83}.
The helicity dependence was measured as the ratio of the helicity-dependent cross section to the p-wave resonance cross section, referred to as longitudinal asymmetry, which amounts to (9.56 $\pm$ 0.35)\%.
The large P-violation was explained as the interference between the amplitudes of the p-wave resonance and the neighboring s-wave resonance~\cite{sus82,sus82E}.
Longitudinal asymmetry was intensively studied in neutron transmission and in (n,$\gamma)$ measurements~\cite{LANL89,LANL91,mas89,shi93}.
The $\gamma$-ray energy dependence of the asymmetry was not found, which implies that the interference occurs in the entrance channel to the compound state and not in the exit channel~\cite{shimem}.
Under this assumption, the longitudinal asymmetry $A_{\rm L}$ is given by
\begin{equation}
A_{\rm L} \simeq - \frac{2xW}{E_{\rm p}-E_{\rm s}}\sqrt{\frac{\Gamma_{\rm s}^{\rm n}}{\Gamma_{\rm p}^{\rm n}}}
\label{eq:Asy}
\end{equation}
assuming s-wave and p-wave resonances, where $E_{\rm s}$ and $E_{\rm p}$ are their respective energies, $\Gamma_{\rm s}^{\rm n}$ and $\Gamma_{\rm p}^{\rm n}$ are the corresponding neutron widths and $W$ is the weak matrix element.
The value of $x$ is defined as $x^2 = \Gamma_{{\mathrm p},j=\frac12}^{\rm n}/\Gamma^{\rm n}_{\rm p}$,
where $\Gamma_{{\rm p},j=\frac12}^{\rm n}$ is the partial neutron width for the total angular momentum of the incident neutron $j=1/2$. The detail definition of $x$ is described in Appendix E.
A theoretical model explaining the large enhancement of P-violation in compound states was studied and summarized in Ref.~\cite{mit01}.
The enhancement mechanism is expected to be applicable to P- and T-violating interactions and to enable highly sensitive explorations of CP-violating interactions beyond the Standard Model of elementary particles. The sensitivity of T-violation can be quantified in relation to the magnitude of P-violation as a function of $x$~\cite{gud92,gud17}.
\par However, the values of $x$ and $W$ have not yet been measured individually. The value of $x$ can be extracted from the energy dependence of the angular distribution of $\gamma$-rays from a p-wave resonance in a neutron induced compound nucleus, which has not yet been measured. Theoretically, it can be deduced assuming interference between partial waves in the entrance channel~\cite{fla85}.
\par 
In this paper we report measurement results of the angular distribution of individual $\gamma$-rays emitted from 0.74~eV p-wave resonance of $^{139}$La+n as a function of incident neutron energy.

\section{Experiment}
\subsection{Experimental Setup}
\begin{figure}
 \begin{center}
 \includegraphics[width=0.9\linewidth]{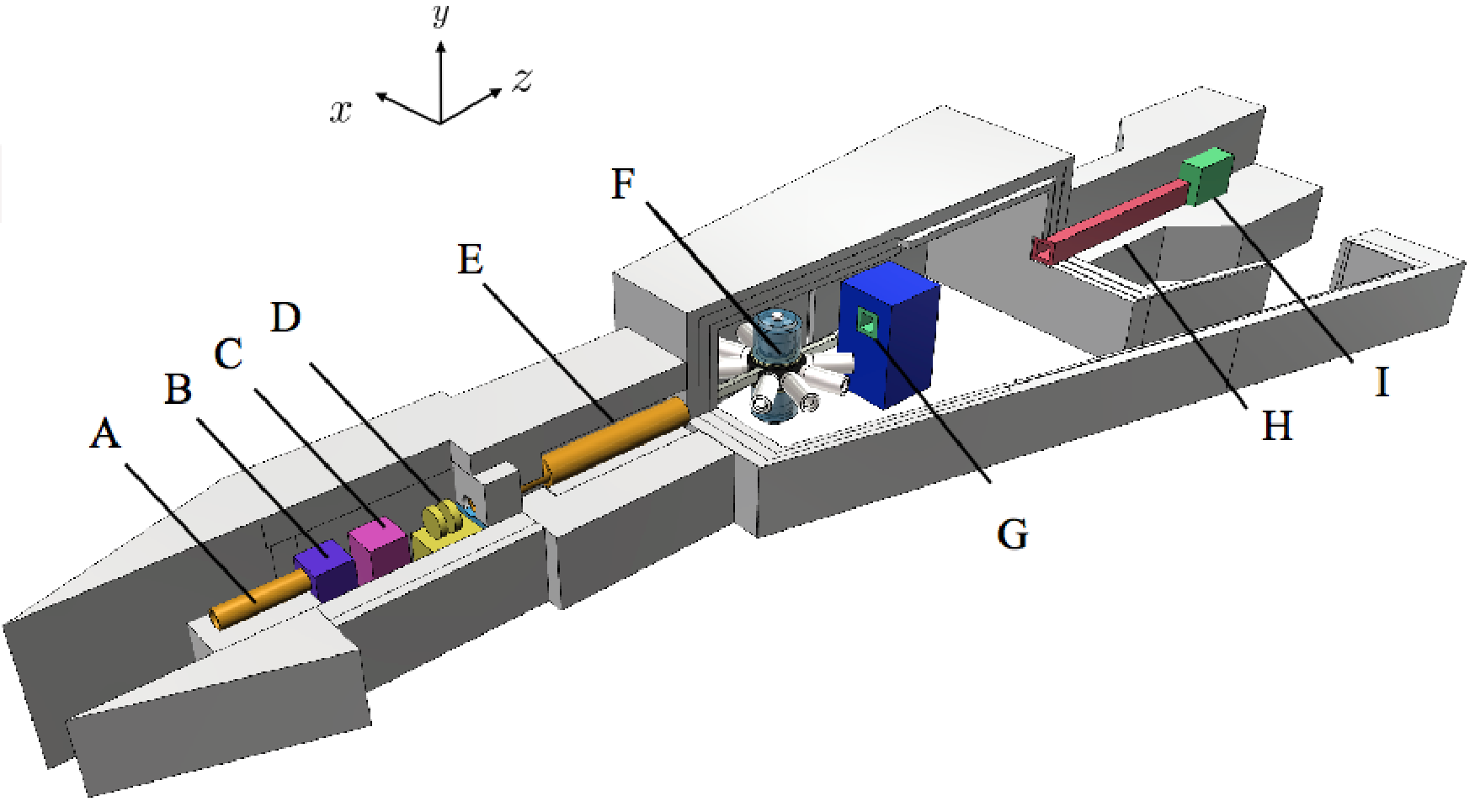}

 \caption{Schematic of the ANNRI installed at the beamline 04 of MLF at J-PARC.
 	(A) Collimator,
	(B) T0-chopper,
	(C) Neutron filter,
	(D) Disk chopper,
	(E) Collimator,
	(F) Germanium detector assembly,
	(G) Collimator,
	(H) Boron resin,
	and
	(I) Beam stopper (Iron).
}
 \label{ANNRI}
 \end{center}
\end{figure}

The angular distribution of individual $\gamma$-rays through the radiative capture reactions induced by epithermal neutrons was measured by introducing a pulsed neutron beam into the Accurate Neutron--Nucleus Reaction Measurement Instrument (ANNRI) installed at the beamline BL04 of the Material and Life science experimental Facility (MLF) of the Japan Proton Accelerator Research Complex (J-PARC), as shown in Fig.~\ref{ANNRI}~\cite{ANNRI}.
The primary proton beam pulses were injected to the neutron production target in a single-bunch mode with a repetition rate of 25 Hz and an average beam power of 150 kW during the measurement.
The disk chopper was operated synchronously with the proton injection for the suppression of low energy neutrons, to avoid frame overlap.
The beam collimation was adjusted to define the neutron beam in a $22$ mm diameter circle on the target, placed at $21.5$ m from the moderator surface~\cite{kin11}.
A lead plate (thickness: $37.5$ mm) was placed in the upstream optics to suppress the $\gamma$-ray background.
The $z$-axis is defined in the beam direction, the $y$-axis is the vertical upward axis, and $x$-axis is perpendicular to them, thus $xyz$ forms a right-handed frame.

\begin{figure}[htbp]
	\centering
	\includegraphics[width=0.9\linewidth]{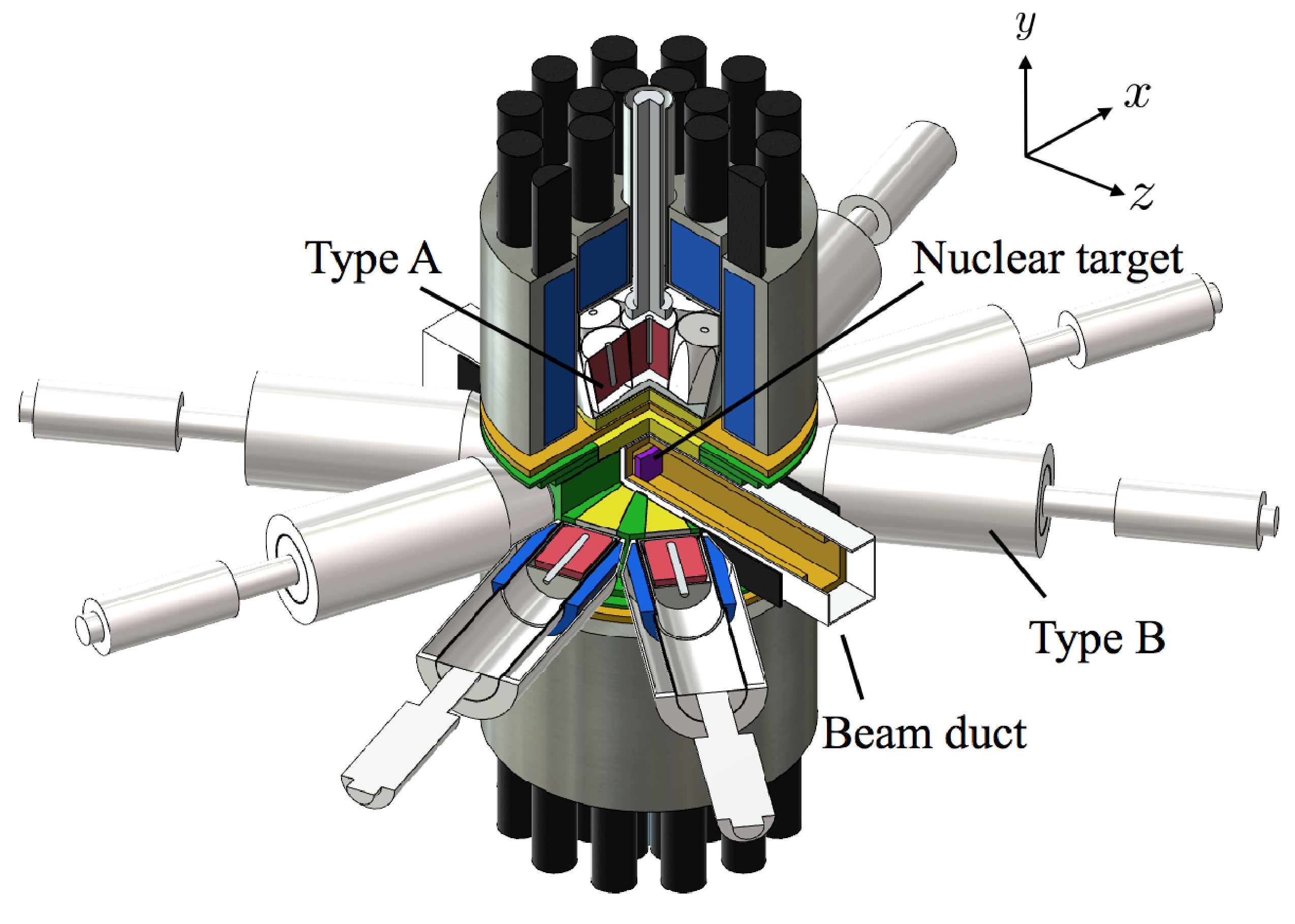}
	\caption[]{Configuration of the germanium spectrometer assembly.}
	\label{Ge}
\end{figure}

An assembled set of high-purity germanium spectrometers were used to detect $\gamma$-rays emitted from the target~\cite{kim12}.
The configuration of the germanium spectrometer assembly is shown in Fig.~\ref{Ge}.

\begin{figure}[htbp]
	\centering
	\includegraphics[width=0.9\linewidth]{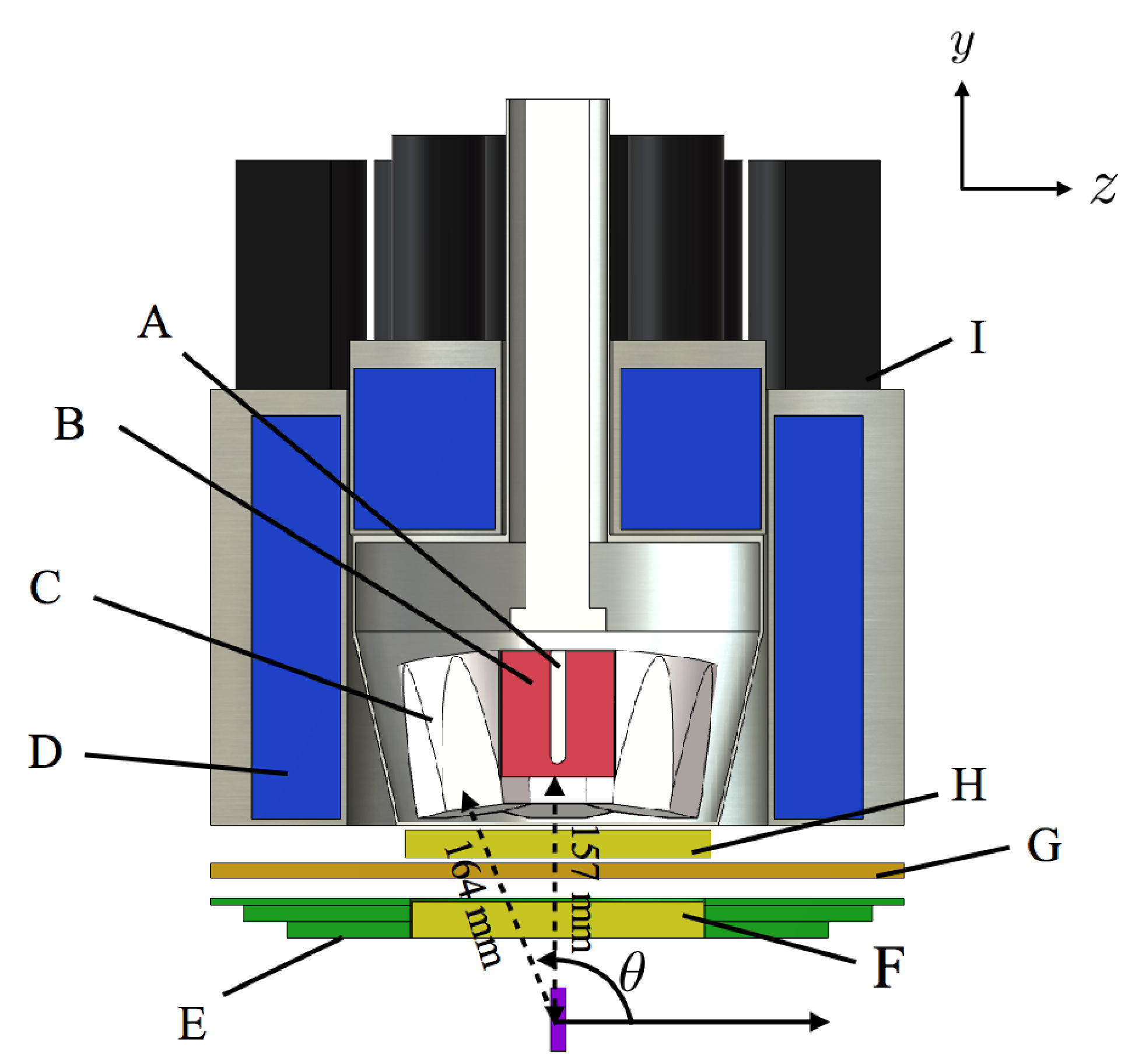}
	\caption[]{
		Schematics of the upper seven Type-A detectors.
		(A) Electrode,
		(B) Germanium crystal,
		(C) Aluminum case,
		(D) BGO crystal,
		(E) $\gamma$-ray shield (Pb collimator),
		(F) Neutron shield-1 (LiH 22.3 mm thick),
		(G) Neutron shield-2 (LiF 5 mm thick),
		(H) Neutron shield-3 (LiH 17.3 mm thick),
		and
		(I) Photomultiplier tube for BGO crystal.
	}
	\label{Ge1}
\end{figure}

\begin{figure}[htbp]
	\centering
	\includegraphics[width=0.9\linewidth]{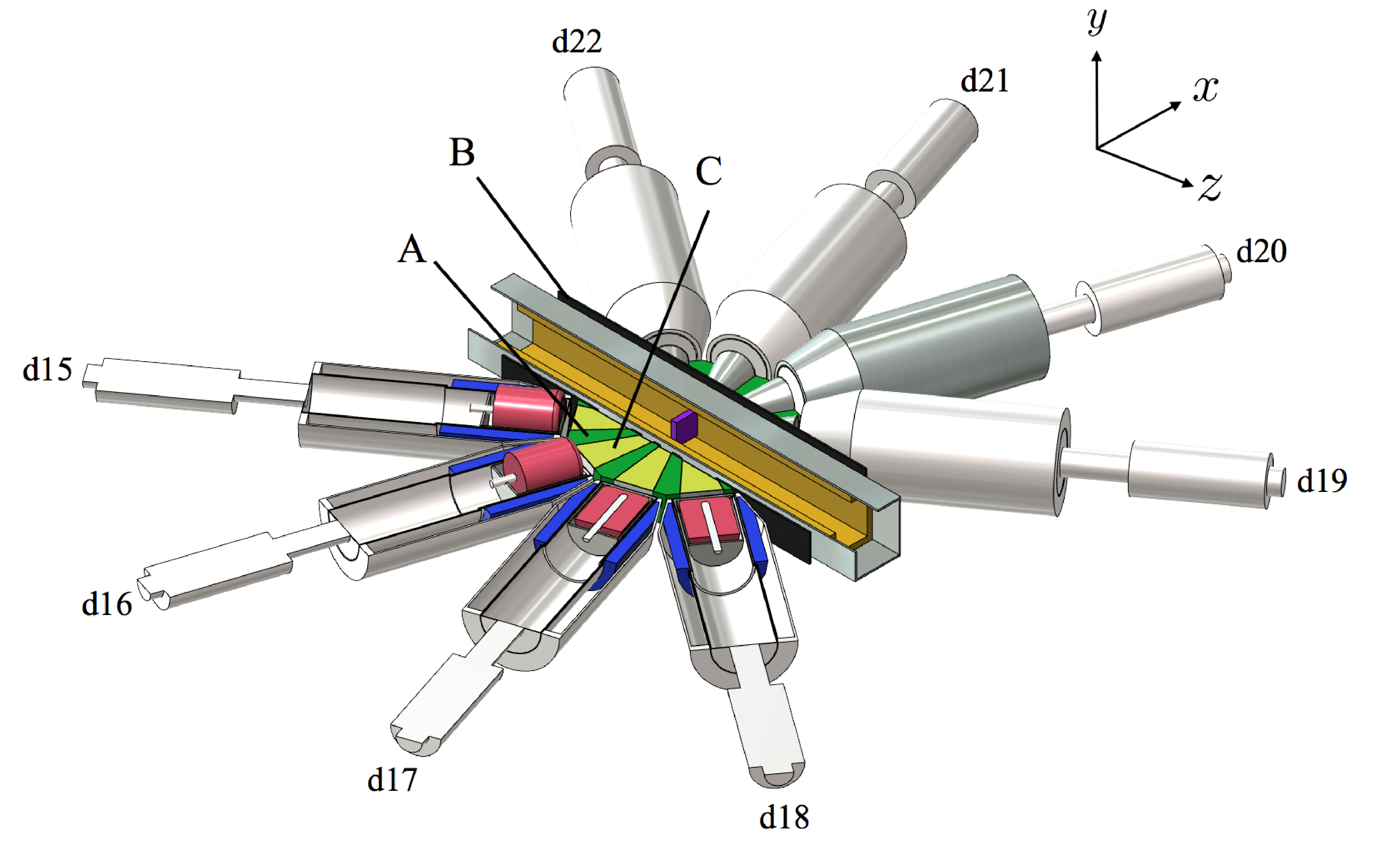}
	\caption[]{Schematics of Type-B detectors.
		(A) Pb collimator,
		(B) Carbon board,
		and
		(C) LiH powder.
	}
	\label{Ge2}
\end{figure}

The assembly consisted of two types of detector units: {Type-A} (Fig.~\ref{Ge1}) and {Type-B} (Fig.~\ref{Ge2}).

Two combined seven Type-A detectors were placed above and below the target. The shape of the Type-A detector was hexagonal to enable clustering as shown in Fig.~\ref{Ge1}.
The polar angles between the center of the target and the center of the crystal surface facing the target were $\theta = 71^{\circ}$, $90^{\circ}$, and $109^{\circ}$.

Eight Type-B detectors were placed on the $xz$-plane at $\theta = 36^{\circ}$, $72^{\circ}$, $108^{\circ}$, and $144^{\circ}$, as shown in Fig.~\ref{Ge2}. 
The central crystals of the upper and lower Type-A detectors are denoted as d1 and d8, respectively, and the other surrounding six detectors are denoted d2-d7
(d9-d14). The names of each Type-B detectors are shown in Fig.~\ref{Ge2}. 
In our measurement, d16 and d17 were not used.

All germanium crystals were operated at a temperature of $77$ K. 
The typical energy resolution for the $1.332$ MeV $\gamma$-rays was $4.2$ keV.

\begin{figure}[htbp]
	\centering
	\includegraphics[width=0.9\linewidth]{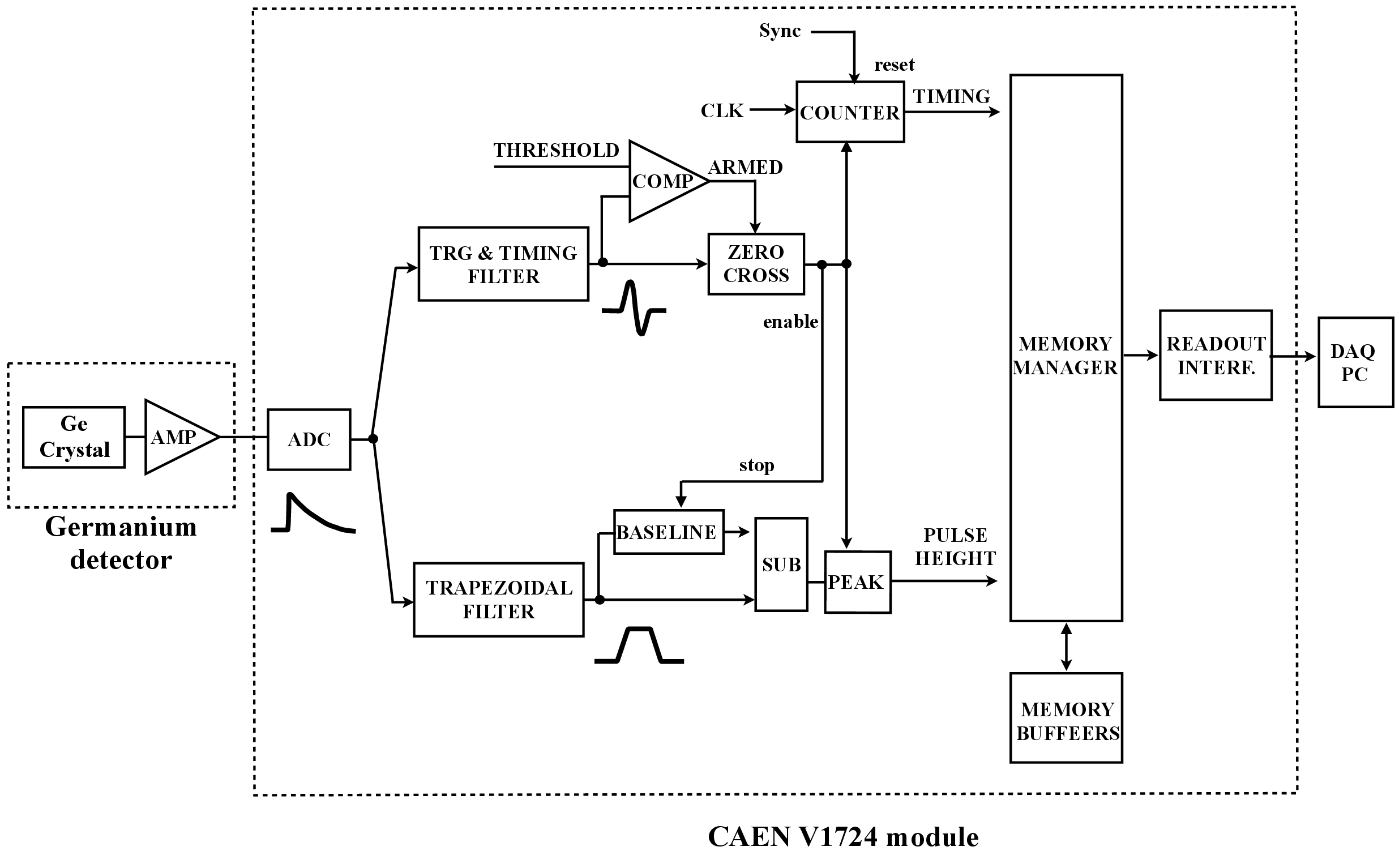}	
	\caption[]{
		Block diagram of the signal processing. A signal from the germanium detector is divided into two branches: one for the timing and triggering and the other for the pulse height. In the branch of timing and triggering, the signal is converted to a bipolar signal. Signals over a threshold are triggered and the time of zero crossing determines the timing of the signal. In the branch of the pulse height, the signal is converted to a trapezoidal signal and the height of the trapezoidal from a baseline determines the pulse height of the signal.
	}
	\label{DAQ}
\end{figure}

The output signal from each crystal was processed independently. The block diagram of the signal processing is shown in Fig.~\ref{DAQ}.
The output signals from the preamplifier were fed into the signal processing module CAEN V1724~\cite{CAEN}, which stored the combination of the pulse height digitized using the peak-sensitive ADC and the timing of the zero-cross point measured from the timing pulse of the injection of the primary proton beam bunch $t^{\rm m}$.
The CAEN V1724 module transferred the stored data to the computer when 1024 event data are accumulated in the local buffer.
Two pulses temporarily closer than $0.4 ~\mu$s were processed as a single event, while their pulse heights were recorded as a zero when their time difference was in the range of $0.4~\mu$s to $3.2~\mu$s.

Their response functions were simulated using GEANT4.9.6.  Definitions of the symbols to describe the detector characteristics and results of the simulation are discussed in detail in Appendix A.

The relation between the pulse height of photo peaks and the deposit $\gamma$-ray energy was determined by observing $\gamma$-rays emitted in neutron capture reactions by aluminum. 
The effective photo-peak efficiency including the solid angle coverage of each detector unit was determined relatively based on the assumption that prompt $\gamma$-rays from $^{14}$N($\rm{n}, \gamma)$ of a melamine target were emitted isotropically. The relative photo-peak efficiencies are shown in Table~\ref{tab:DetectorP}.

\subsection{Measurement}
\begin{figure}[htbp]
	\centering
	\includegraphics[width=0.9\linewidth]{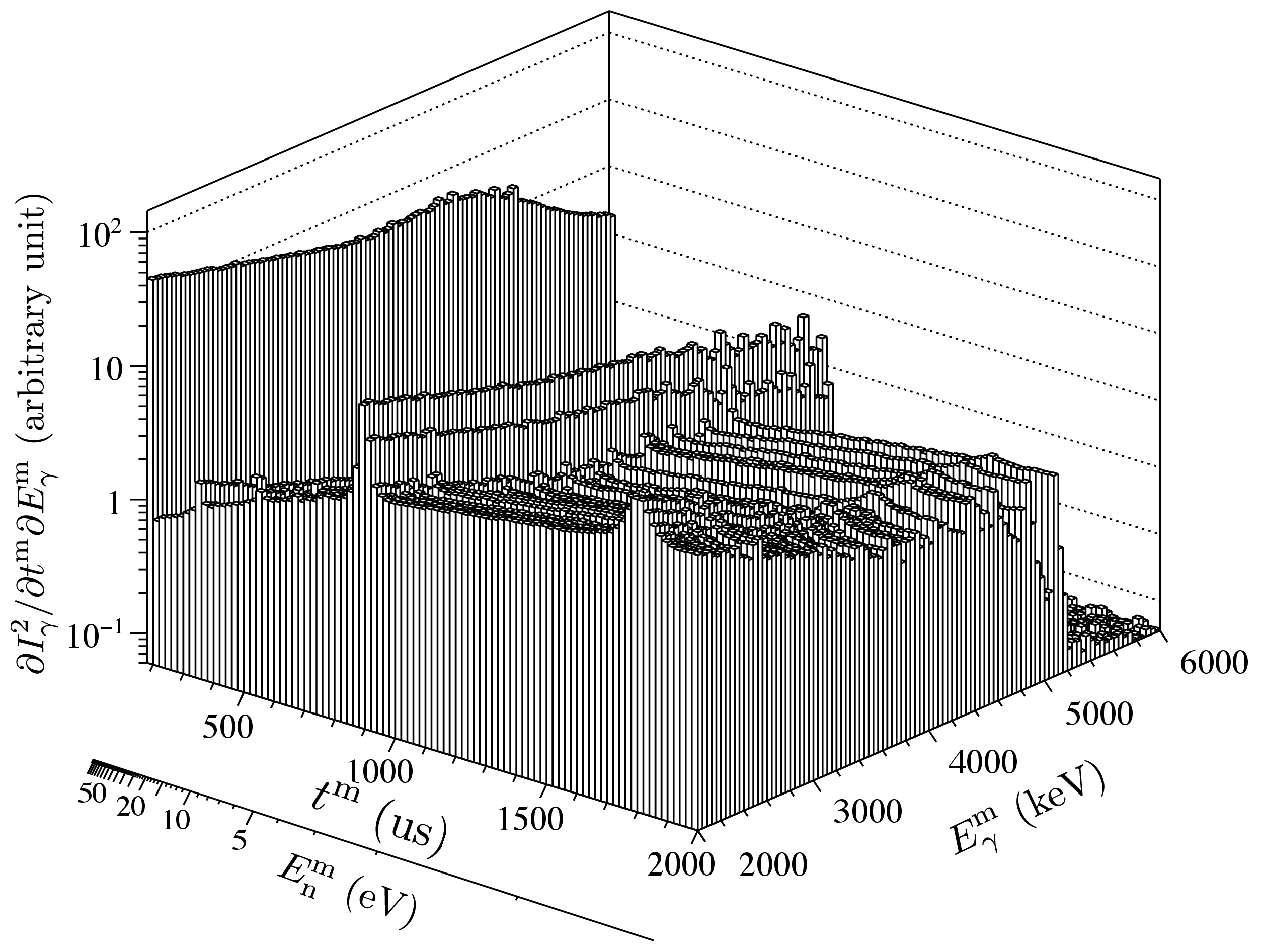}
	\caption[]{
	$\partial^2I_\gamma/\partial t^{\rm m}\partial E^{\rm m}_\gamma$ 2-dimensional histogram of $\gamma$-rays with the lanthanum target as a function of timing $t^{\rm m}$ and deposit $\gamma$-ray energy $E^{\rm m}_\gamma$.
	The corresponding neutron energy $E^{\rm m}_{\rm n}$ is also shown.
			}
	\label{TOFGamma1}
\end{figure}

\begin{figure}[htbp]
	\centering
		\includegraphics[width=0.9\linewidth]{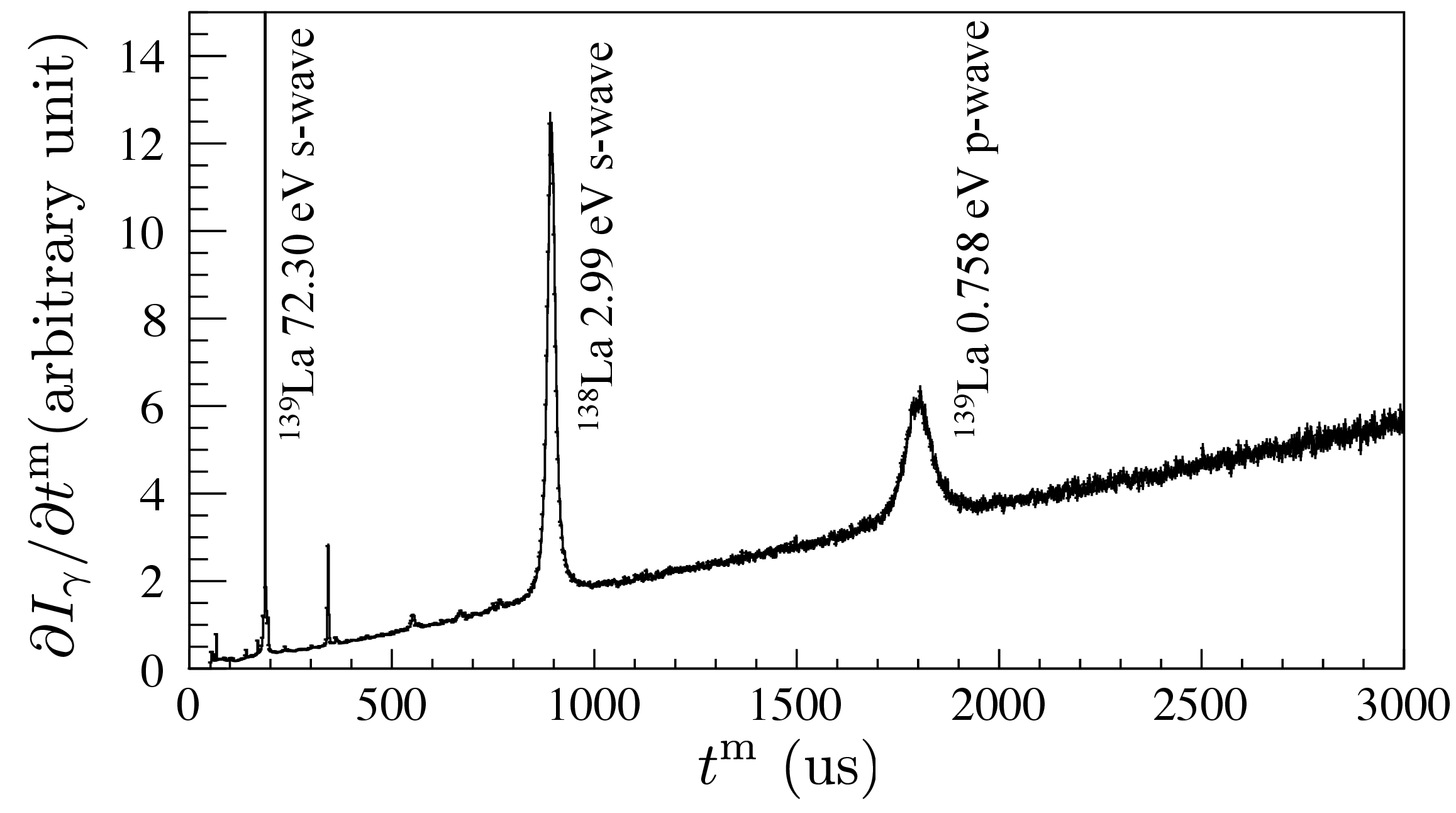}
	\caption[]{
	$\gamma$-ray counts relatively corrected by the incident beam intensity as a function of $t^{\rm m}$ for $E^{\rm m}_\gamma \ge 2\,{\rm MeV}$, which is referred as $\partial I_{\gamma} / \partial t^{\rm m}$.
	}
	\label{TOF}
\end{figure}

\begin{figure}[htbp]
	\centering
	\includegraphics[width=0.9\linewidth]{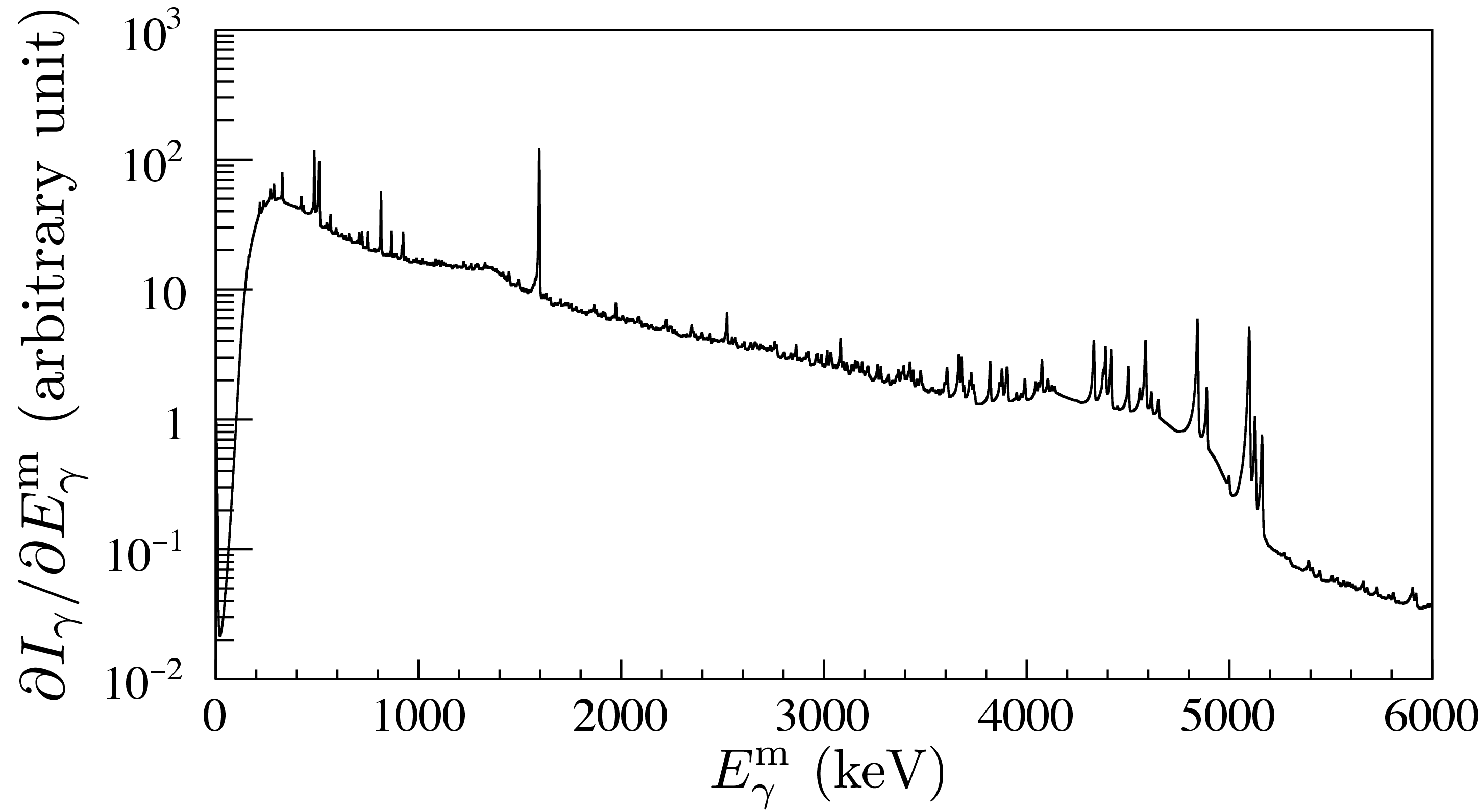}
	\caption[]{
	Pulse height spectrum of $\gamma$-rays $\partial I_{\gamma} / \partial E^{\rm m}_\gamma$ from the (n,$\gamma$) reaction with lanthanum target as a function of $E^{\rm m}_\gamma$.
	}
	\label{Gamma1}
\end{figure}

The target was a natural-abundance lanthanum plate at room temperature, with the dimensions of $40\,{\rm mm}\times 40\,{\rm mm}\times 1\,{\rm mm}$, and with a purity of $99.9$\%.
The total number of $\gamma$-ray events detected in the experiment are denoted $I_{\gamma}$.
Here, the corresponding neutron energy $E^{\rm m}_{\rm n}$ is defined as
\begin{eqnarray}
E^{\rm m}_{\rm n} = \frac{m_{\rm{n}}}{2}\left(\frac{L}{t^{\rm m}}\right)^2 ,
\end{eqnarray}
where $m_{\rm{n}}$ is the neutron mass and $L$ is the distance between the target and the moderator surface.
The deposited $\gamma$-ray energy $E^{\rm m}_\gamma$ obtained from the calibration of the pulse height is defined as well.
The obtained results are shown as a 2-dimensional histogram corresponding to $\partial^2I_{\gamma}/\partial t^{\rm m}\partial E^{\rm m}_\gamma$ in Fig.~\ref{TOFGamma1}.
The histograms projected on $t^{\rm m}$ and $E^{\rm m}_\gamma$ are shown in Fig.~\ref{TOF} and Fig.~\ref{Gamma1}, respectively.
In Fig.~\ref{TOF}, $\gamma$-ray events with $E^{\rm m}_\gamma$ are integrated and relatively corrected by the incident beam spectrum for $t^{\rm m}$.
The incident beam spectrum was obtained by measuring the 477.6~keV $\gamma$-rays in ${}^{10}$B(n,$\alpha \gamma)^{7}$Li reactions, with a boron target placed at the detector center.
The small peak at $t^{\rm m}\sim1800~\mu\rm{s}$ is a p-wave resonance, and the $1/v$ component is the tail of an s-wave resonance in the negative energy region as listed in Table~\ref{tab:Igamma-fit}.

The neutron energy in the center-of-mass system $E_{\rm n}$ is given as
\begin{eqnarray}
E_{\rm n}=\frac{m_{\rm n}m_{\rm A}}{m_{\rm n}+m_{\rm A}}
	\left(\frac{\bm{p}_{\rm n}}{m_{\rm n}}-\frac{\bm{p}_{\rm A}}{m_{\rm A}}\right)^2
,
\end{eqnarray}
where $\bm{p}_{\rm n}$ is the momentum of the neutron in the laboratory system, $\bm{p}_{\rm A}$ is nuclear momentum of the target, and $m_{\rm A}$ is the mass of the nucleus of the target.

\begin{figure}[htbp]
	\centering
	\includegraphics[width=0.8\linewidth]{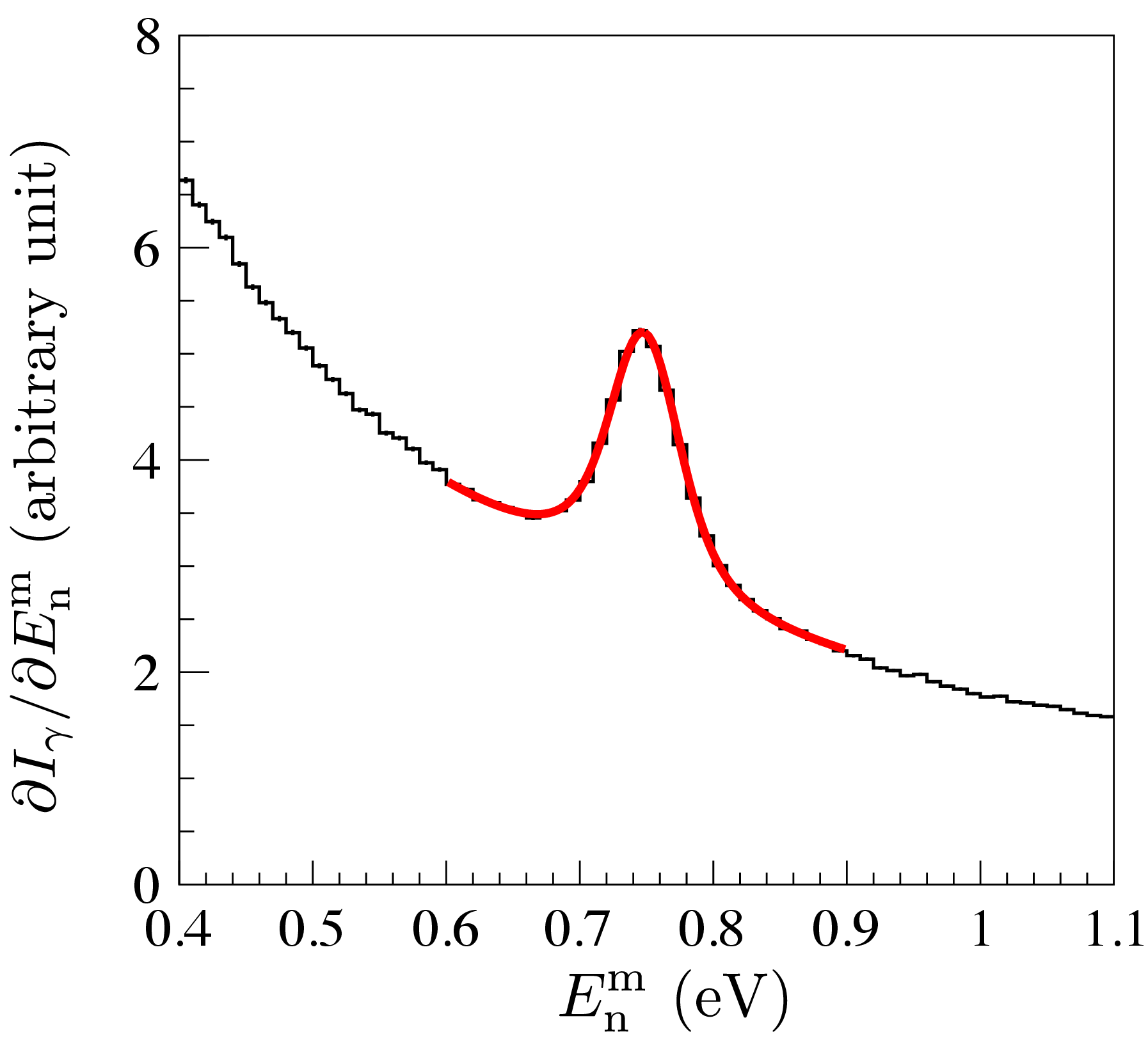}
	\caption[]{
	Fitted result of the p-wave resonance. The curve shows the best fit.
	}
	\label{fig:Igamma-fit}
\end{figure}

The beam divergence is sufficiently small and the following can be assumed: 
\begin{eqnarray}
\bm{p}_{\rm n} = \sqrt{2m_{\rm n}E_{\rm n}} \, \bm{e}_z
,
\end{eqnarray}
where $\bm{e}_z$ is the unit vector parallel to the beam axis.
The resonance energy $E_{r}$ and the total width $\Gamma_{r}$ of the $r$-th resonance, which are obtained by fitting $\partial I_{\gamma}/\partial t^{\rm m}$ with Eq.~\ref{eq:Igamma}  are shown in Table~\ref{tab:Igamma-fit}, together with the published values. The formalism of the neutron absorption cross section is described in Appendix B. The pulse shape of the neutron beam and the Doppler effect of the target nucleus are considered, as shown in Appendix C and Appendix D, respectively. As the neutron width of the p-wave resonance is negligibly smaller than $\gamma$-ray width of the p-wave resonance, the total width of p-wave resonance was used as the $\gamma$-ray width of p-wave resonance. 
The fitted result is shown in Fig.~\ref{fig:Igamma-fit}.

\begin{table*}[htbp]
\begin{center}
	\begin{tabular}{c||c|c|c|c|c|c||c|c}
	\multirow{2}{*}{$r$} & 
	\multicolumn{6}{c||}{published values} &
	\multicolumn{2}{c}{this work} \\
	\cline{2-9}
	&
	$E_r\,[{\rm eV}]$ & 
	$J_r$ & 
	$l_r$ & 
	$\Gamma^{\gamma}_{ r}\,[{\rm meV}]$ &
	$g_{ r}\Gamma^{n}_{ r}\,[{\rm meV}]$&
	 $g_{ r}\Gamma_r^{{\rm n}l_r}\,[{\rm meV}]$ &
	$E_r\,[{\rm eV}]$ & 
	$\Gamma^{\gamma}_{ r}\,[{\rm meV}]$ \\
	\hline
	$1$ & $-48.63$$^{\rm(a)}$ & $4$$^{\rm(a)}$ & $0$ &$62.2^{\rm(a)}$ & & $82^{\rm(a)}$ & \\
	$2$ & $0.758\pm0.001$$^{\rm(b)}$ & & $1$ & $40.11\pm1.94$$^{\rm(c)}$ &  $(5.6 \pm0.5 )\times 10^{-5}$$^{\rm(c)}$& &
		$0.740\pm 0.002$ & $40.41\pm 0.76$ \\
	$3$ & $72.30\pm0.05$$^{\rm(b)}$ & & $0$ & $75.64\pm2.21$$^{\rm(c)}$ &$11.76\pm0.53$$^{\rm(c)}$& &
		$$ & $$ \\
	\end{tabular}
	\caption{
	Resonance parameters of $^{139}$La.
	(a) taken from Ref.~\cite{mughabghab} and Ref.~\cite{JENDLall}.
	(b) taken from Ref.~\cite{JENDL-La-28}.
	(c) calculated from Refs.~\cite{JENDL-La-26} and \cite{JENDL-La-28}.
	$\Gamma_r^{{\rm n}l_r}$ is a reduced neutron width.
	}
	\label{tab:Igamma-fit}
\end{center}	
\end{table*}

\begin{figure}[htbp]
	\centering
	\includegraphics[width=0.9\linewidth]{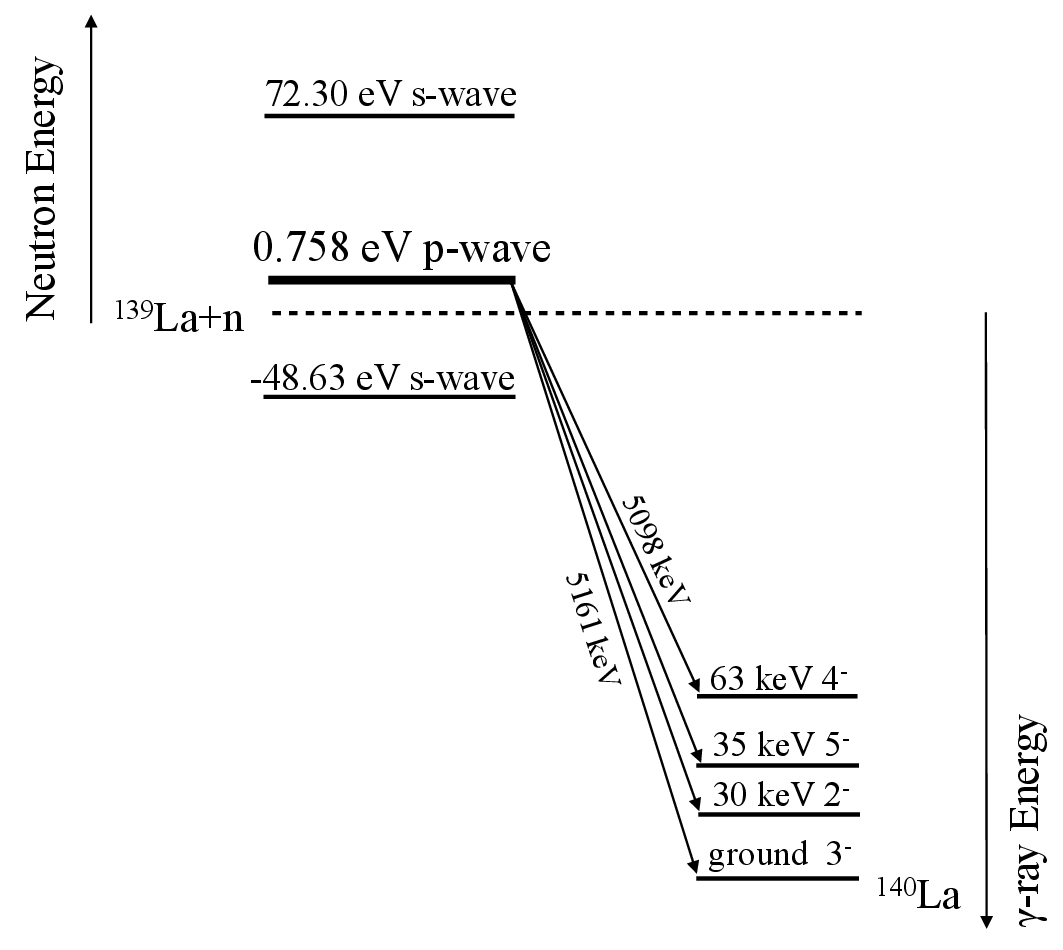}
	\caption[]{
	Transitions from $^{139}$La+n to $^{140}$La. Dashed line shows separation energy of $^{139}$La+n.
	}
	\label{TransitionLa}
\end{figure}

\begin{figure}[htbp]
	\centering
	\includegraphics[width=0.9\linewidth]{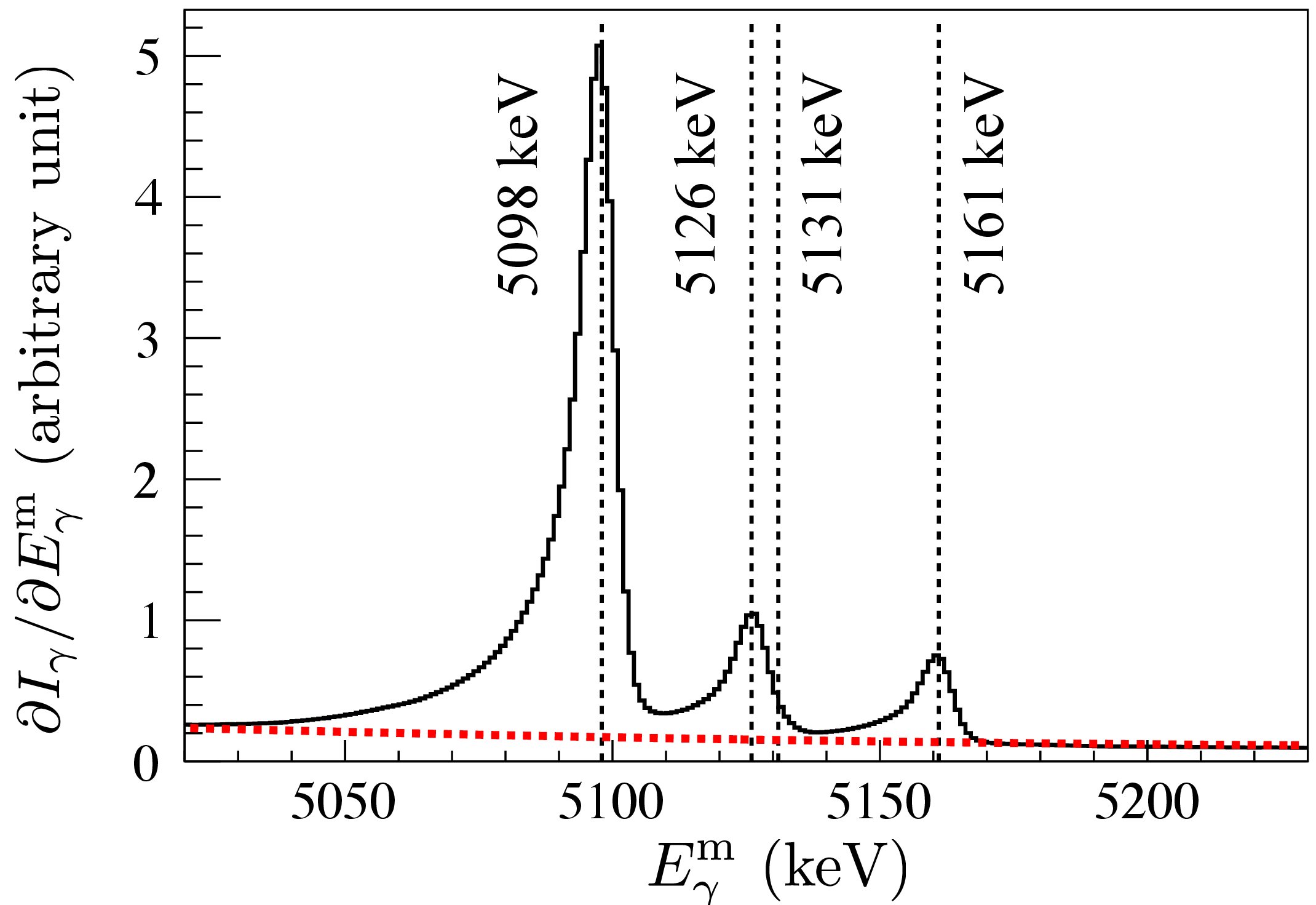}
	\caption[]{
	Expanded $\partial I_{\gamma}/\partial E_\gamma^{\rm m}$. The dotted line shows the background determined by the simulation.}
	\label{Gamma2}
\end{figure}

The level scheme related to $^{139}$La(n,$\gamma$)${}^{140}$La reaction is schematically shown in Fig.~\ref{TransitionLa}~\cite{NNDC}.
The $\gamma$-ray transitions to the ground state and low excited states of ${}^{140}$La were observed as shown in the expanded $\partial I_{\gamma}/\partial E^{\rm m}_\gamma$ (Fig.~\ref{Gamma2}).
The highest peak at $E^{\rm m}_\gamma$$=$$5161\,{\rm keV}$ corresponds to the $\gamma$-ray direct transition to the ground state of ${}^{140}$La (spin of the final state:$F$=3), the middle peak corresponds to the overlap of two transitions at $E^{\rm m}_\gamma$$=$$5131\,{\rm keV}$ and $5126\,{\rm keV}$ to the first and second excited states at excited energy $30\,{\rm keV}~(F\rm{=5}), 35\,{\rm keV}~($$F$=2), and the lower peak at $E^{\rm m}_\gamma$$=$$5098\,{\rm keV}$ corresponds to the excited state at $63\,{\rm keV}~(F$=4).

\begin{figure}[htbp]
	\centering
	\includegraphics[width=0.9\linewidth]{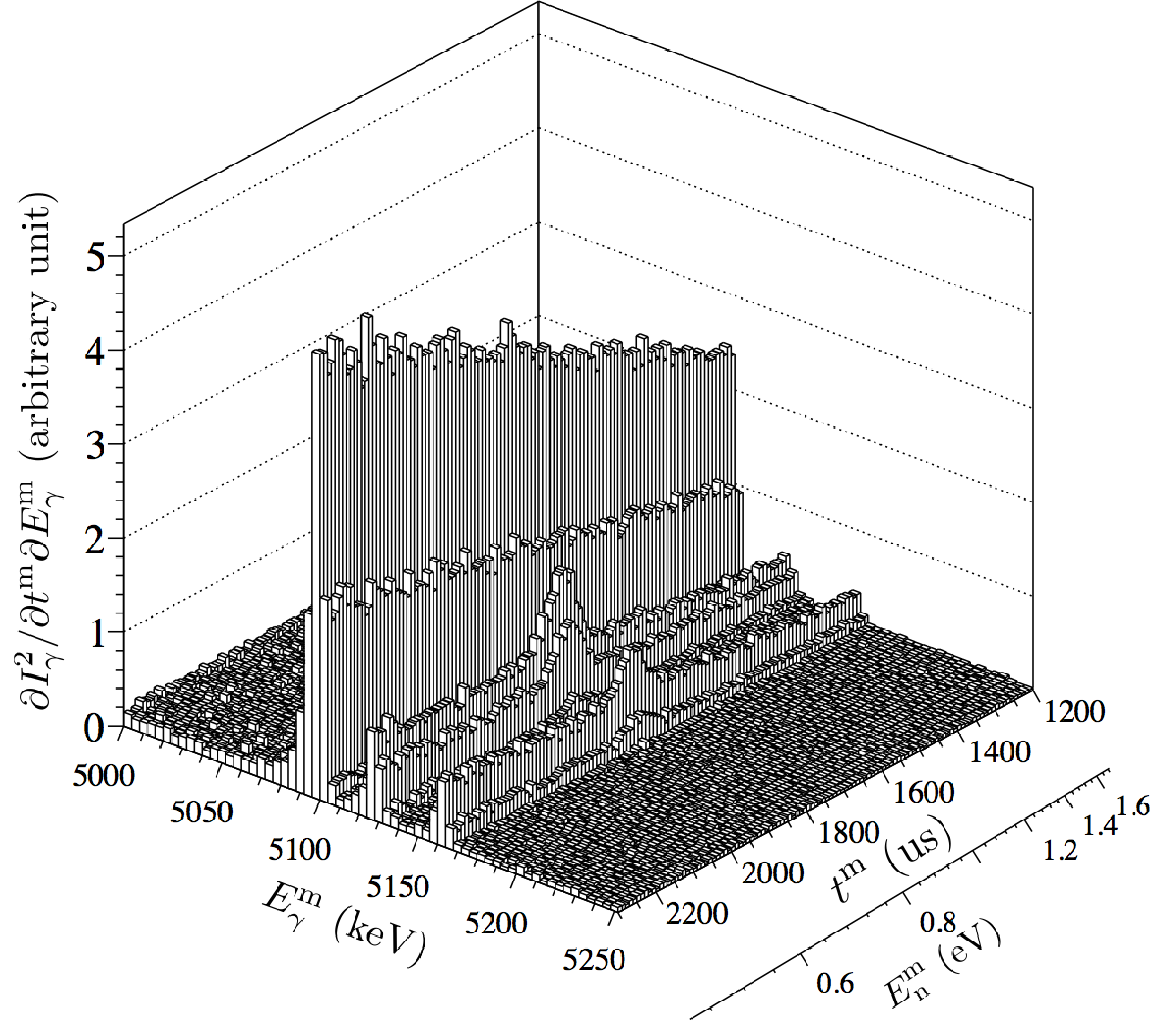}
	\caption[]{
	Magnified 2-dimensional histogram.
	}
	\label{TOFGamma2}
\end{figure}

Figure~\ref{TOFGamma2} shows the magnified 2-dimensional histogram of $\partial^2 I_{\gamma}/\partial t^{\rm m} \partial E^{\rm m}_\gamma$ in the vicinity of the p-wave resonance and the $\gamma$-ray transition to the ground state of ${}^{140}$La.
The p-wave resonance was selectively observed only for two ridges corresponding to the transition at $E^{\rm m}_\gamma$$=$$5161\,{\rm keV}$ and the sum of transitions at $E^{\rm m}_\gamma$$=$$5131\,{\rm keV},5126\,{\rm keV}$, but not for the ridge at $E^{\rm m}_\gamma$$=$$5098\,{\rm keV}$.
According to the dependence of $\partial^2 I_{\gamma}/\partial t^{\rm m} \partial E^{\rm m}_\gamma$ on $t^{\rm m}$, and therefore on the incident neutron energy, the s-wave resonance in the negative region contributes to all three $\gamma$-ray transitions, and the p-wave resonance contributes to the $5161\,{\rm keV}$ transition and $5131\,{\rm keV}$ and/or $5126\,{\rm keV}$ transitions.

\subsection{Angular Distribution}
Hereafter, we concentrate on the $5161\,{\rm keV}$ transition to the ground state of ${}^{140}$La, in order to study the interference between the s- and p-wave amplitudes.

The photo-peak efficiency, including both the detection efficiency and the solid angle coverage, was readjusted using the photo-peak counts of the $\gamma$-rays at $E^{\rm m}_\gamma$$=$$5262\,{\rm keV}$ from the ${}^{14}$N(n,$\gamma)$ reaction measured using the melamine target.
It can be reasonably assumed that the $\gamma$-rays are emitted isotropically, as the ${}^{14}$N does not have any resonance below $400\,{\rm eV}$ and p-wave or higher angular momentum components of the incident neutron is negligibly small in this energy region.

The photo-peak counts of the $5161\,{\rm keV}$ transition were determined by subtracting the background counts caused by Compton scattering of the more energetic $\gamma$-rays from targets other than the lanthanum target.
To evaluate the background, two energy regions were used: (I) $5200\,{\rm keV}$$\le$$E^{\rm m}_\gamma$$\le$$5290\,{\rm keV}$ and (II) $4900\,{\rm keV}$$\le$$E^{\rm m}_\gamma$$\le$$4980\,{\rm keV}$.
The contribution of Compton scattering of $\gamma$-rays corresponding to the three photo-peaks are contained in region (II).
The amount of this contribution from Compton scattering is estimated using the response function $\bar{\psi}$ given in Eq.~\ref{eq:psibar} and obtained by simulation.
The background in region (II) is estimated by subtracting the Compton contribution from the $\gamma$-ray counts in region (II).
The background is estimated using a best-fit third-order polynomial of $E^{\rm m}_\gamma$ in regions (I) and (II).

There still remains a possible contamination of prompt $\gamma$-rays from impurities overlapping with the $5161\,{\rm keV}$ photo-peak.
The possible contamination was examined over the entire pulse height spectrum, and was determined to be less than $0.08\%$ of the photo-peak.
The possibility of contamination was neglected as the determined upper limit of $0.08\%$ is smaller than the statistical error of the photo-peak.

According to the data acquisition system, two pulses detected within $3.2~\mu$s did not have amplitude information, which amounted to $2\%$ of the total $\gamma$-ray counts in the vicinity of the p-wave resonance.
The $2\%$ loss was corrected in the following analysis.

Two pulses detected within $0.4~\mu{\rm s}$ were processed as a single pulse.
The corresponding loss of the events were estimated as $0.2\%$ of the total $\gamma$-ray counts in the vicinity of the p-wave resonance, which is negligibly small compared with the statistical error of the corresponding $\gamma$-ray counts, and is ignored in the following analysis.

\begin{figure*}[t]
	\begin{center}
	\includegraphics[width=140mm]{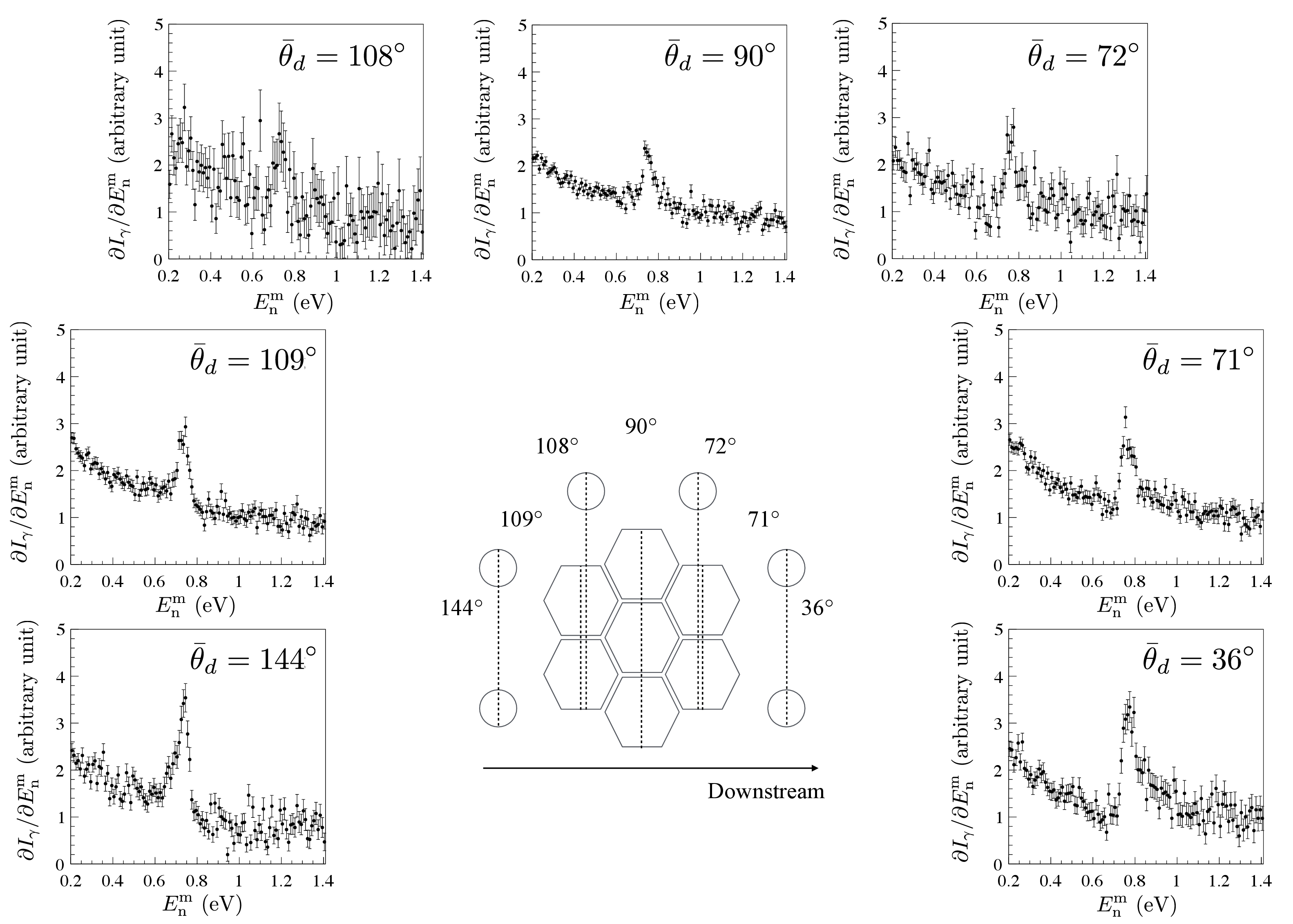}
	\caption[]{
	$\partial N/\partial t^{\rm m}$ in the vicinity of p-wave resonance for each $\bar\theta_d$. The central figure shows degrees in the direction of neutron momentum of the type-A detectors and the type-B detectors. The hexagons and the circles in the center of the figure denote each crystal of the type-A detector and the type-B detector, respectively.
	}
	\label{angular}
	\end{center}
\end{figure*}

Equation~\ref{eq:Igamma} was extended to describe the angular distribution of $\gamma$-rays as
\begin{eqnarray}
&&
\frac{\partial^2 I_{\gamma}}{\partial t^{\rm m}\partial \Omega_{\gamma}} (t^{\rm m},\Omega_{\gamma})
	\nonumber\\
	&& \quad=
	I_0 \int
	{\rm d}E'
	{\rm d}^3p_{\rm A}
	\Phi(t^{\rm m},E',\bm{p}_{\rm A})
	\diff{\sigma_{{\rm n}\gamma}}{\Omega_{\gamma}}(E,\Omega_\gamma)
	,
	\nonumber\\
&&
\Phi(t^{\rm m},E',\bm{p}_{\rm A})
	\nonumber\\
	&& \quad=
	\frac{\partial^2 \phi}{\partial E_{\rm n} \partial t}
		\left(E',t^{\rm m} - L \sqrt{\frac{m_{\rm n}}{2E'}}\right)
	\nonumber\\ && \quad \quad \times
	\frac{1}{(2\pi m_{\rm A}k_{\rm B}T)^{3/2}} e^{-p_{\rm A}^2/2m_{\rm A}k_{\rm B}T}
	\nonumber\\ && \quad \quad \times
	\frac{1}{\sigma_{\rm t}(E)}
	\left(1 - e^{-n\sigma_{\rm t}(E)\,\Delta z}\right)
	.
	\nonumber\\
\label{eq:IgammaOmegaAnother}
\end{eqnarray}
The $\gamma$-ray counts to be measured by the $d$-th detector can be written as
\begin{eqnarray}
	\pdiff{N}{t^{\rm m}}(t^{\rm m},\bar\theta_d)
	&=&
		\int_{\Omega_d} {\rm d}\Omega_{\gamma}
		\int_{(E^{\rm m}_{\gamma})_d^{{w^{-}}}}^{(E^{\rm m}_{\gamma})_d^{{w^{+}}}} {\rm d}(E^{\rm m}_{\gamma})_d
	\nonumber\\&&\times
		\frac{\partial^2 I_{\gamma}}{\partial t^{\rm m}\partial \Omega_{\gamma}} (t^{\rm m},\Omega_{\gamma})
		\psi_d(E_\gamma,\Omega_\gamma,(E^{\rm m}_{\gamma})_d)
	,
	\nonumber\\
\label{eq:Ndgamma}
\end{eqnarray}
where the photo-peak region is taken as the full-width at quarter-maximum, which implies $w=1/4$.
Figure~\ref{angular} shows the $N(t^{\rm m},\bar\theta_d)$ for $5161\,{\rm keV}$ $\gamma$-rays.
The peak shape of the p-wave resonance varies according to $\bar\theta_d$.
Here, we define $N_{\rm L}$ and $N_{\rm H}$ as
\begin{eqnarray}
N_{\rm L}(\theta_\gamma) &=& \int_{E_{\rm p}-2\Gamma_{\rm p}}^{E_{\rm p}}
	\pdiff{N}{t^{\rm m}}(t',\bar\theta_\gamma) {\rm d}t^{\rm m}\frac{dt^{\rm m}}{dE_{\rm n}}dE_{\rm n}
	,
	\nonumber\\
N_{\rm H}(\theta_\gamma) &=& \int_{E_{\rm p}}^{E_{\rm p}+2\Gamma_{\rm p}}
	\pdiff{N}{t^{\rm m}}(t',\bar\theta_\gamma) {\rm d}t^{\rm m}\frac{dt^{\rm m}}{dE_{\rm n}}dE_{\rm n}
	.
\end{eqnarray}
\begin{figure}[hbtp]
	\centering
	\includegraphics[width=0.7\linewidth]{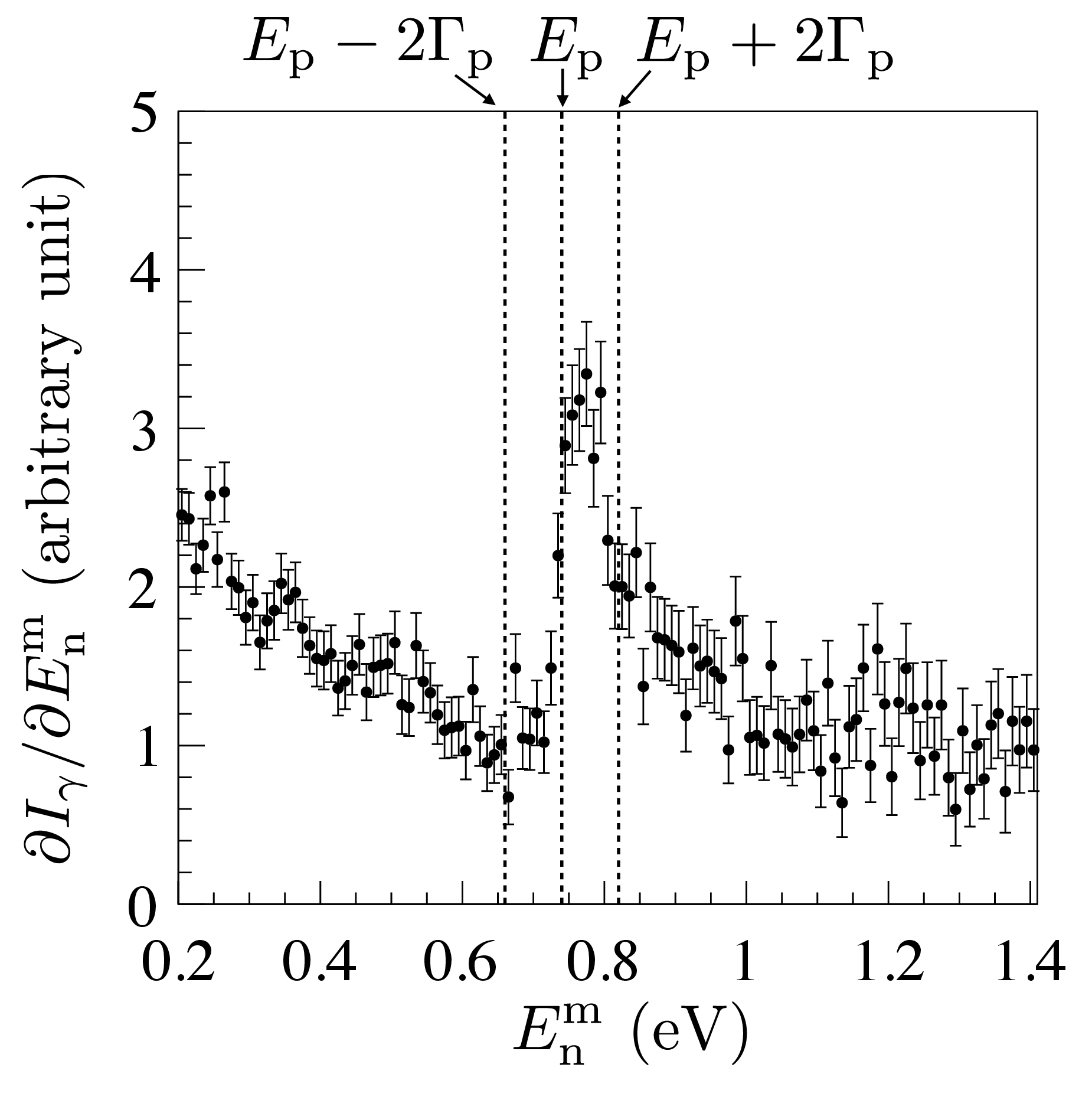}
	\caption[]{
	Visualization of the definition of $N_{\rm L}$ and $N_{\rm H}$.
	}
	\label{nLnHdef}
\end{figure}
The angular dependences of $N_{\rm L}$ and $N_{\rm H}$ are shown in Fig.~\ref{nLandnH}.
\begin{figure}[t]
	\begin{center}
	\includegraphics[width=0.9\linewidth]{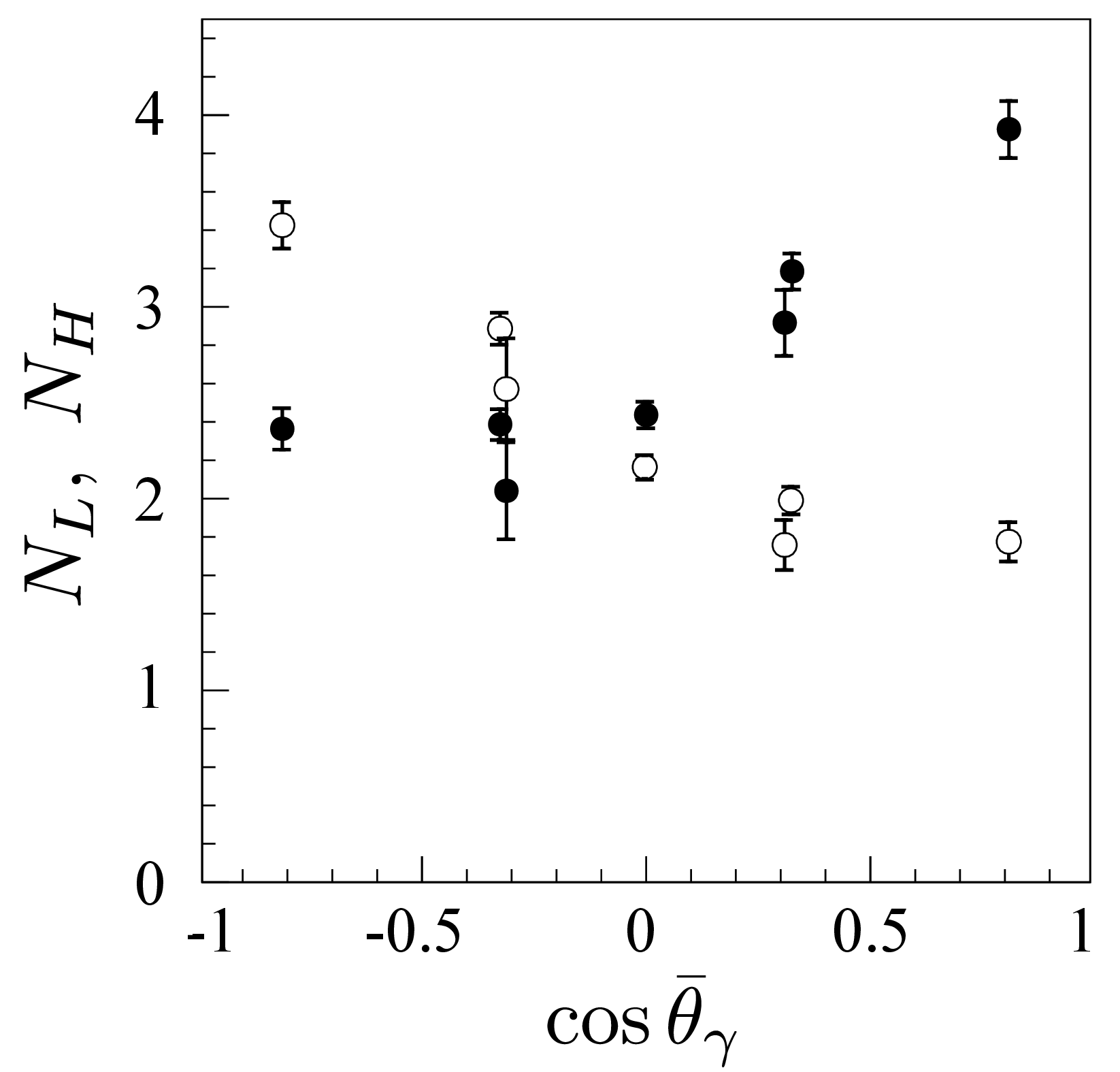}
	\caption[]{
	Angular dependences of $N_{\rm L}$ and $N_{\rm H}$. The white point and black point show $N_{\rm L}$ and $N_{\rm H}$, respectively. 
	}
	\label{nLandnH}
	\end{center}
\end{figure}
As $N_{\rm L}$ and $N_{\rm H}$ have certain angular dependence in Fig.~\ref{nLandnH}, we define an asymmetry between $N_{\rm L}$ and $N_{\rm H}$ to determine the angular dependence of the shape of p-wave resonance as
\begin{eqnarray}
A_{\rm LH} = \asym{N_{\rm L}}{N_{\rm H}}
.
\label{eq:ALH}
\end{eqnarray}
The angular dependence of $A_{\rm LH}$ is shown in Fig.~\ref{Asy3}. 
\begin{figure}[t]
	\begin{center}
	\includegraphics[width=0.9\linewidth]{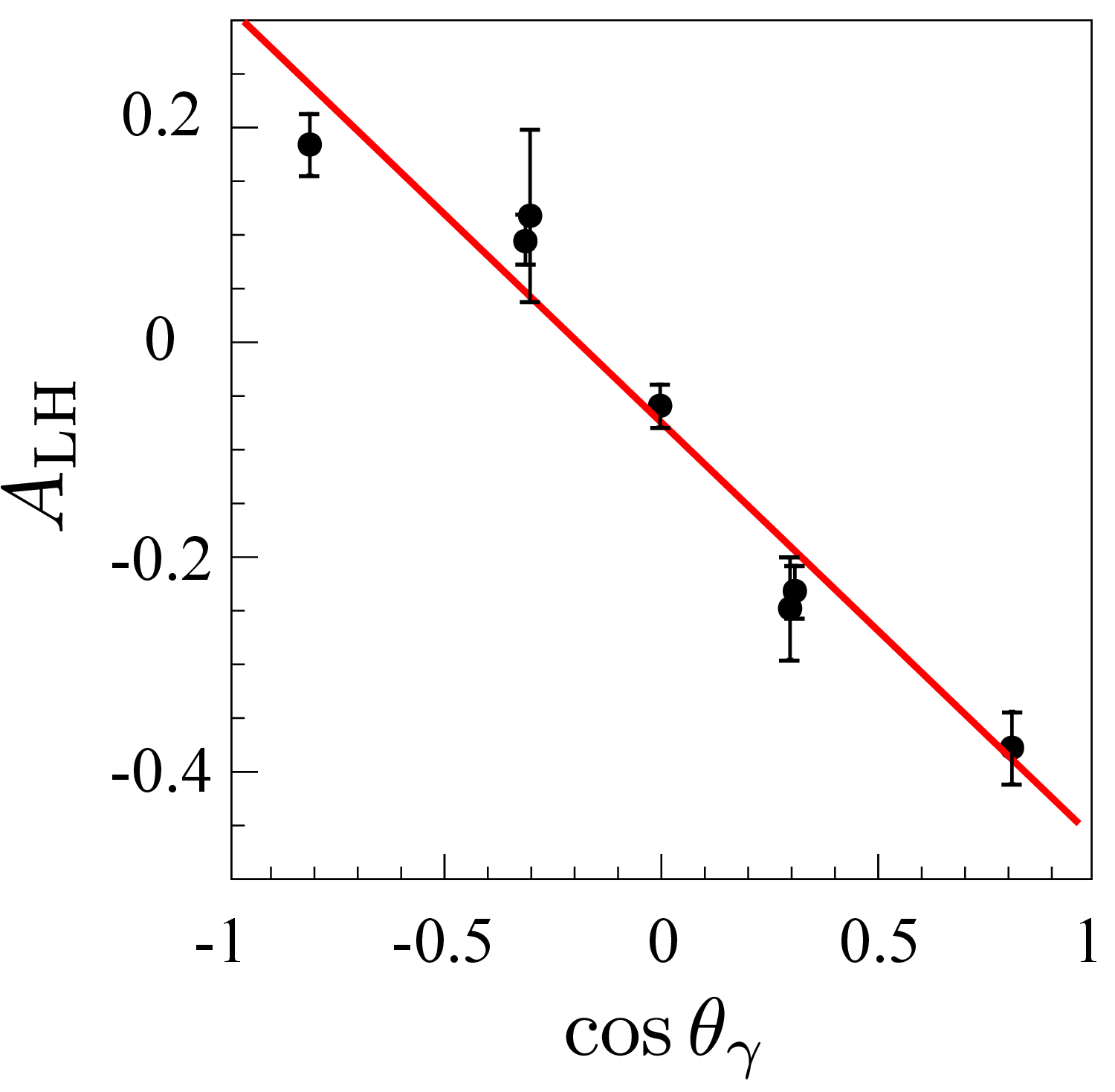}
	\caption[]{
	Angular dependence of $A_{\rm LH}$. The solid line shows the best fit. 
	}
	\label{Asy3}
	\end{center}
\end{figure}
The asymmetry $A_{\rm LH}$ has a correlation with $\cos\bar\theta_\gamma$ as
\begin{eqnarray}
	A_{\rm LH} = (A \cos\bar\theta_\gamma+B)
\end{eqnarray}
where
\begin{eqnarray}
A = -0.3881\pm 0.0236,\ B = -0.0747\pm 0.0105,
\label{eq::ALH_value}
\end{eqnarray}
which are the best fit results of Fig.~\ref{Asy3}. This result implies that clear energy dependence of the angular distribution of $\gamma$-rays was observed.\\

The $x$ value of the p-wave resonance can be obtained using this result. The analysis and the interpretation of this result are discussed in section \ref{sec:analysis} and \ref{sec:discussion}.
\section{Analysis}\label{sec:analysis}
Our experimental results are analyzed using the formulation of possible angular correlations of individual $\gamma$-rays, emitted in (n,$\gamma$) reactions induced by low energy neutrons according to s- and p-wave amplitudes~\cite{fla85}. The formalism of the differential cross section of the (n,$\gamma$) reaction induced by unpolarized neutrons is described in Appendix E.
We use $\bm{I}$ as the spin of the target nuclei, $\bm{J}$ as the spin of the compound nucleus, $\bm{F}$ as the spin of the final state of the $\gamma$-ray transition, and $\bm{l}$ as the orbital angular momentum of the incident neutron.
The total neutron spin is defined as $\bm{j}=\bm{l}+\bm{s}$, where $\bm{s}$ is the neutron spin.
The value of $j$ is $1/2$ for s-wave neutrons ($l$$=$$0$) and $j$$=$$1/2$, $3/2$ for p-wave neutrons ($l$$=$$1$).

\begin{figure}[hbtp]
	\centering
	\includegraphics[width=0.9\linewidth]{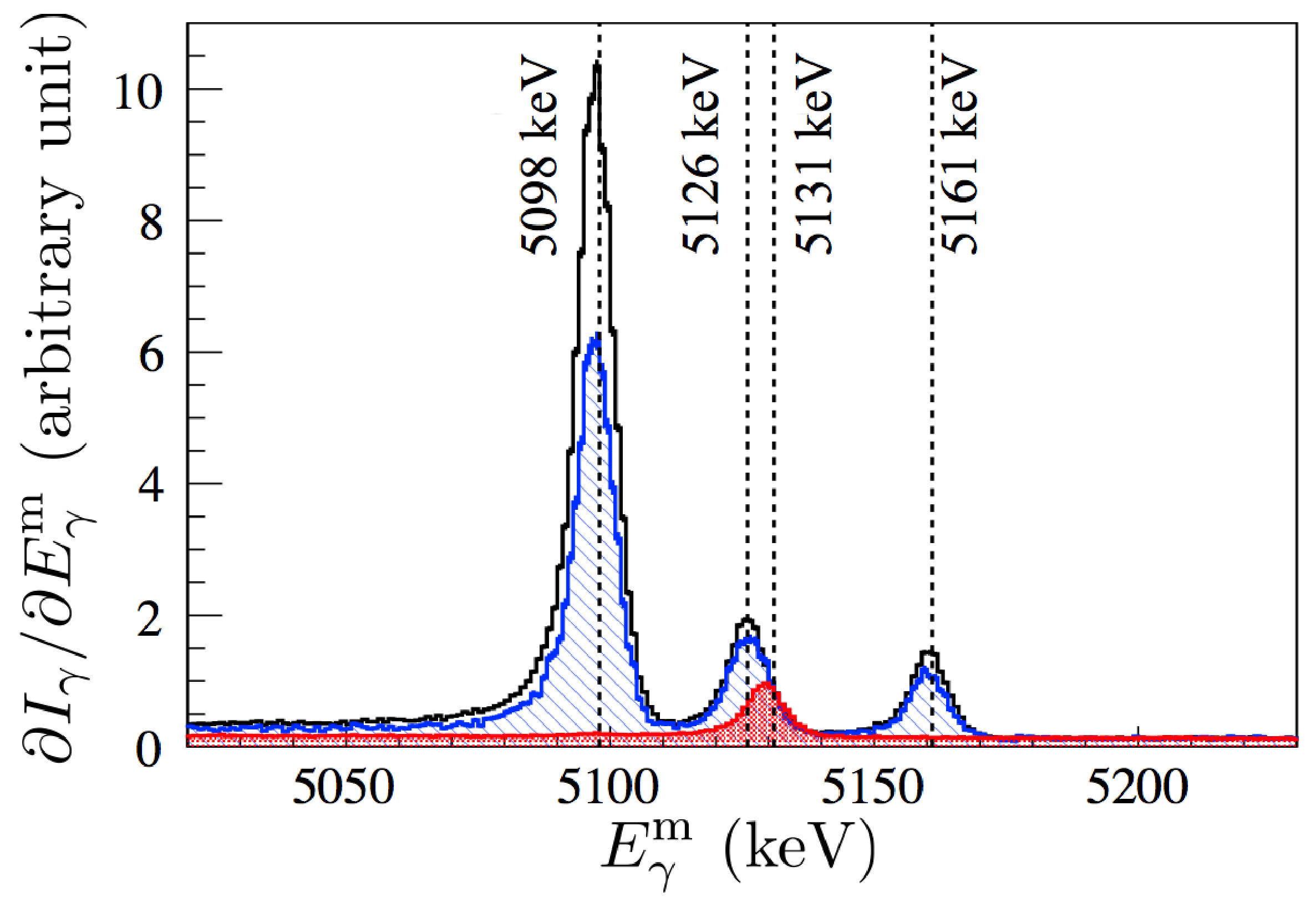}
	\caption[]{
	Comparison of the expanded $\partial I_{\gamma}/\partial E^{\rm m}_{\gamma}$ gated in the vicinities of s$_{1}$-wave resonance ($E^{\rm m}_{\rm n}$=0.2-0.4~eV: black line), the p-wave resonance ($E^{\rm m}_{\rm n}$=0.6-0.9~eV: shaded area with diagonal line), and the s$_{2}$-wave resonance ($E^{\rm m}_{\rm n}$=70-75~eV : solid shaded area).
	}
	\label{fig:Igamma72eV}
\end{figure}

The p-wave resonance and two neighboring s-wave resonances are considered in the negative and positive energy region, listed in Table~\ref{tab:res}, in the following analysis.
\begin{table}[htbp]
\begin{center}
	\begin{tabular}{c||c|c|c|c|c||c}
	$r$ &
	$E_r\,[{\rm eV}]$ & 
	$J_r$ & 
	$l_r$ & 
	$\Gamma^{\gamma}_r\,[{\rm meV}]$ &
	$g_{ r}\Gamma^{\rm{n}}_r\,[{\rm meV}]$ &
	\\
	\hline
	$1$ & $-48.63^{\rm (a)}$ & $4^{\rm (a)}$ & $0$ &$62.2^{\rm (a)}$ &$(571.8)^{\rm (a) \ast}$ &${\rm s}_1$\\
	$2$ & $0.740\pm 0.002$ & $4$ & $1$ & $40.41\pm 0.76$ & $(5.6 \pm0.5 )\times 10^{-5}$$^{\rm (c)}$ & ${\rm p}$\\
	$3$ & $72.30\pm0.05^{\rm (b)}$ & $3$ & $0$ & $75.8\pm5.4^{\rm (c)}$ & $11.76\pm0.53^{\rm (c)}$ & ${\rm s}_2$\\
	\end{tabular}
	\caption{
	Resonance parameters of $^{139}$La used in the analysis. (a) taken from Ref.~\cite{mughabghab} and Ref.~\cite{JENDLall}.
	(b) taken from Ref.~\cite{JENDL-La-28}.
	(c) calculated from Refs.~\cite{JENDL-La-26} and \cite{JENDL-La-28}.
	$^{\ast}$The neutron width for the negative resonance was calculated using $\abs{E_1}$ instead of $E_1$.
	}
	\label{tab:res}
\end{center}	
\end{table}
The resonance energy and resonance width measured in this work is adopted to the p-wave resonance ($r$$=$$2$) and the values in Ref.~\cite{mughabghab} for the negative resonance and positive s-wave resonance($r$$=$$1$, $r$$=$$3$).
The compound nuclear spin of the negative s-wave resonance is $J_1$$=$$4$~\cite{mughabghab}.
The compound nuclear spin of the p-wave resonance is assumed to be the same as that of the negative s-wave resonance, as both the negative s-wave and the p-wave components were observed in the $5161\,{\rm keV}$ $\gamma$-ray transition, which implies that $J_2$$=$$4$.
Nevertheless, the compound nuclear spin of the positive s-wave resonance is taken as $J_3$$=$$3$, as the $5161\,{\rm keV}$ transition was not adequately
observed in the resonance, as shown in Fig.~\ref{fig:Igamma72eV}.
The contributions of far s-wave resonances are assumed to be negligibly small, that is, $\alpha_1=0$ in Eq.~\ref{eq::flambaum2}.

 The ratios of the $\gamma$ width from each resonance to the ground state can be determined by a comparison of the peak height ratio of the neutron resonance gated in the 5161 keV photo-peak between s$_1$-wave, p-wave, and s$_2$-wave as  
\begin{eqnarray}
	\frac{\Gamma_{\rm{s_{1}, gnd}}^{\rm\gamma}}{\Gamma_{\rm{s_{1}}}^{\rm\gamma}}: \frac{\Gamma_{\rm{p, gnd}}^{\rm\gamma}}{\Gamma_{\rm{p}}^{\rm\gamma}}: \frac{\Gamma_{\rm{s_{2}, gnd}}^{\rm\gamma}}{\Gamma_{\rm{s_{2},gnd}}^{\rm\gamma}}=1 : 0.796\pm0.020: 0.009\pm0.006. \nonumber 
	\\
	\label{eq:Branch}
\end{eqnarray}
As shown in Fig.~\ref{fig:Igamma72eV} and in Eq.~\ref{eq:Branch}, the branching ratio from the s$_{2}$-wave resonance to the ground state is very small.
We define $(\overline{a_{0}})_{\rm L}$, $(\overline{a_{1}})_{\rm L}$, $(\overline{a_{3}})_{\rm L}$, $(\overline{a_{0}})_{\rm H}$, $(\overline{a_{1}})_{\rm H}$, and $(\overline{a_{3}})_{\rm H}$ as
\begin{eqnarray}
(\overline{a}_{0,1,3})_{\rm L} &=& 
	\int_{E_{\rm p}-2\Gamma_{\rm p}}^{E_{\rm p}}
	{\rm d}E'
	\int
	{\rm d}^3p_{\rm A}
	\,
	a_{0,1,3}
	\Phi(t^{\rm m},E',\bm{p}_A)
	,
	\nonumber\\
(\overline{a}_{0,1,3})_{\rm H} &=& 
	\int_{E_{\rm p}}^{E_{\rm p}+2\Gamma_{\rm p}}
	{\rm d}E'
	\int
	{\rm d}^3p_{\rm A}
	\,
	a_{0,1,3}
	\Phi(t^{\rm m},E',\bm{p}_A)
	.
	\nonumber\\
\label{eq:a_overline}
\end{eqnarray}

\begin{figure}[hbtp]
	\centering
	\includegraphics[width=0.9\linewidth]{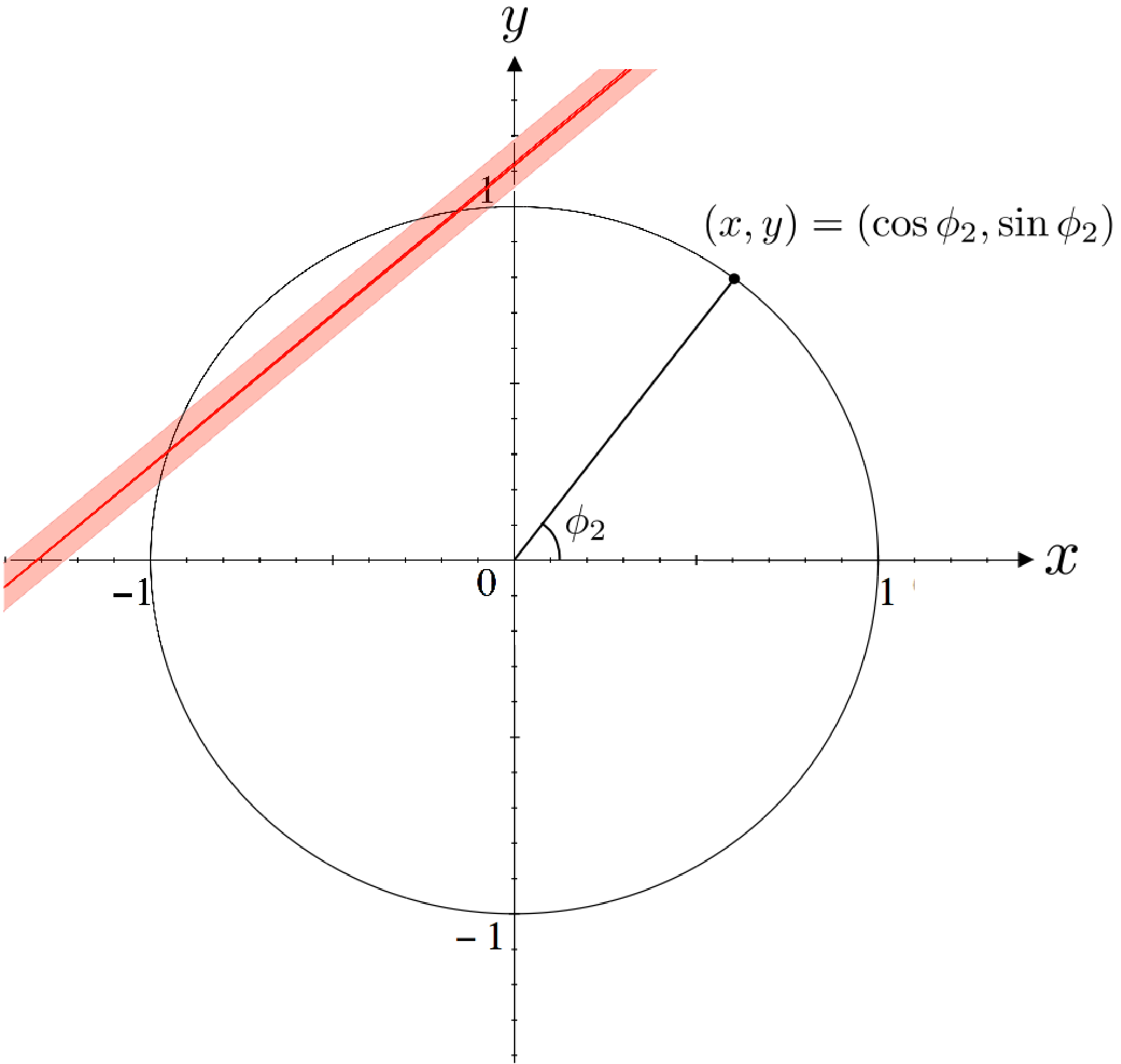}
	\caption[]{
	Visualization of the value of $\phi_2$ on the $xy$-plane. The solid line and shaded area show the central values of $\phi_2$ and 1$\sigma$ area, respectively. $\phi_2$ denotes an angle of a point on a unit circle.
	}
	\label{fig:a0a1}
\end{figure}

Here, the $a_3$ term is ignored as it is proportional to ${\lambda_{2f}}^2$ and it is suppressed relative to the s-wave neutron width, according to the centrifugal potential by the factor of $(kR)^2$.
Under this approximation, Eq.~\ref{eq:Flambaum1} is reduced to
\begin{eqnarray}
\diff{\sigma_{{\rm n}\gamma_f}}{\Omega_\gamma}
	&=& \frac{1}{2}\left(
	a_{0}+a_{1}\cos\theta_{\gamma}
	\right)
	,
	\nonumber\\
	\label{eq:Flambaum2}
\end{eqnarray}
Substituting Eq.~\ref{eq:Flambaum2} into Eq.~\ref{eq:IgammaOmegaAnother}, the angular dependence of the $\gamma$-ray counts in the neutron energy regions $E_{\rm p}-2\Gamma_{\rm p} \le E_{\rm{n}} \le E_{\rm p}$ and $E_{\rm p}\le E_{\rm{n}} \le E_{\rm p}+2\Gamma_{\rm p}$ can be written as
\begin{eqnarray}
\left(\frac{\partial^2 I_{\gamma}}{\partial t^{\rm m}\partial \Omega_{\gamma}} (t^{\rm m},\Omega_{\gamma})\right)_{\rm L}
	&=&
	\frac{I_0}{2} \left( (\overline{a_0})_{\rm L} + (\overline{a_1})_{\rm L} P_1(\cos\theta_\gamma) \right),
	\nonumber\\
	\left(\frac{\partial^2 I_{\gamma}}{\partial t^{\rm m}\partial \Omega_{\gamma}} (t^{\rm m},\Omega_{\gamma})\right)_{\rm H}
	&=&
	\frac{I_0}{2} \left( (\overline{a_0})_{\rm H} + (\overline{a_1})_{\rm H} P_1(\cos\theta_\gamma) \right).
	\nonumber\\
\end{eqnarray}
By convoluting with Eq.~\ref{eq:Ndgamma}, the $\gamma$-ray counts $\left( I_{\gamma,d} \right)_{\rm L}$ and $\left( I_{\gamma,d} \right)_{\rm H}$ to be measured by the $d$-th detector can be written as
\begin{eqnarray}
	\left( I_{\gamma,d} \right)_{\rm L,H}
	&=&
		\frac{I_0}{2} \left(
			(\overline{a_0})_{\rm L,H} \overline{P_{d,0}}
			+
			(\overline{a_1})_{\rm L,H} \overline{P_{d,1}}
		\right)
	.
\end{eqnarray}
As the energy dependence of $x_2$ and $y_2$ is negligibly small in the vicinity of the p-wave resonance ($r_{\rm p}=2$), $(\overline{a_1})_{\rm L}$ and $(\overline{a_1})_{\rm H}$ are linear functions of $x_2$ and $y_2$, thus a function of $\phi_2$.
The value of $\phi_2$ is determined by comparing $\left(\left( I_{\gamma,d} \right)_{\rm L}-\left( I_{\gamma,d} \right)_{\rm H}\right)/\left(\left( I_{\gamma,d} \right)_{\rm L}+\left( I_{\gamma,d} \right)_{\rm H}\right)$ with the measured values $A$ in Eq.~\ref{eq::ALH_value}.

\begin{eqnarray}
A&=&\frac{(\overline{a_1})_{\rm L}-(\overline{a_1})_{\rm H}}{(\overline{a_0})_{\rm L}+(\overline{a_0})_{\rm H}}\nonumber\\
&=&0.295 \cos\phi_2-0.345 \sin\phi_2.
\label{eq:Igamma}
\end{eqnarray}
Two solutions can be obtained as 
\begin{eqnarray}
\phi_2 =
	(99.2_{-5.3}^{+6.3})^{\circ}
	, \quad 
	(161.9_{-6.3}^{+5.3})^{\circ}
	.
	\label{eq:phi}
\end{eqnarray}
The visualization of $\phi_2$ is shown in Fig.~\ref{fig:a0a1}.

\section{Discussion}\label{sec:discussion}
As the value of $\phi_2$ was obtained in the previous section, the T-violation sensitivity is discussed in this section.
We obtain $x_{2}$ from Eq.~\ref{eq:phi}  and Eq.~\ref{eq:xandy} with resonance number $r_{\mathrm p}=2$ as
\begin{eqnarray}
x_{2} =
	-0.16_{-0.11}^{+0.09}
	, \quad 
	-0.95_{-0.03}^{+0.04}
	.\label{eq:xsolution}
\end{eqnarray}
This leads to the value of $W$ which is given by Eq.~\ref{eq:Asy} as 
\begin{eqnarray}
W =
	(13.2_{-5.3}^{+18.1})\ \mathrm{meV}
	,\quad
	 (2.21_{-0.06}^{+0.10})\ \mathrm{meV}
	.\ 
\end{eqnarray}
The published value of $A_{\rm L}=(9.56 \pm 0.35) \times 10^{-2}$ in Ref.~\cite{LANL89} and resonance parameters in Table.~\ref{tab:Igamma-fit} are used in the calculation. Note that the neutron width of the negative s-wave resonance $\Gamma^{\mathrm n}_{\mathrm 1}$ at the resonance energy of the p-wave resonance is adopted.

\begin{figure}[hbtp]
	\centering
		\includegraphics[width=0.9\linewidth]{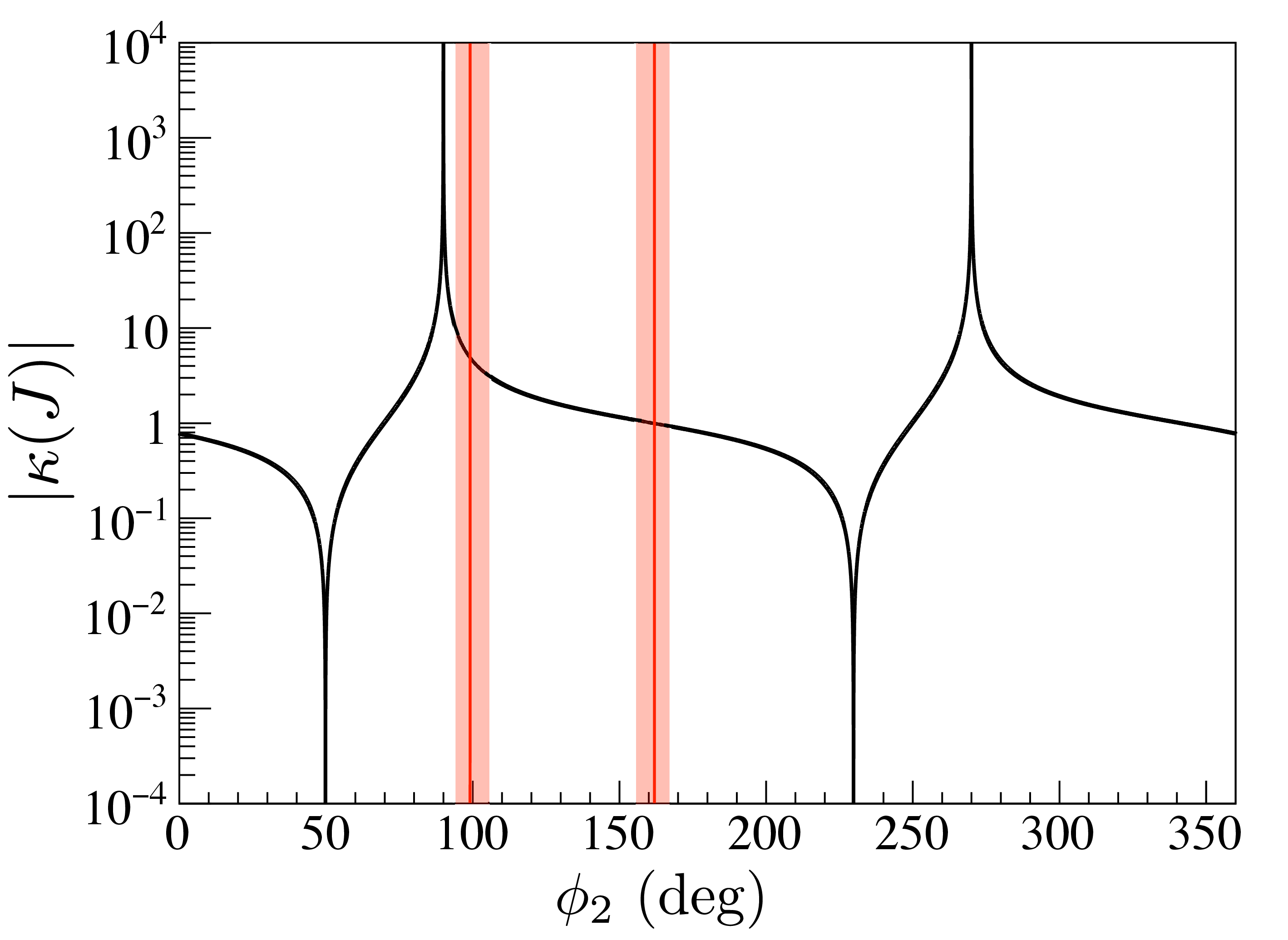}
	\caption[]{
	Value of $\abs{\kappa(J)}$ as a function of $\phi_2$. The solid line and shaded area show the central values of $\phi_2$ and the 1$\sigma$ area from central value, respectively. 
	}
	\label{fig:kappa}
\end{figure}

The ratio of P-odd T-odd cross sections to P-odd cross sections is given as
\begin{eqnarray}
\frac{\Delta\sigma_{\rm PT}}{\Delta\sigma_{\rm P}}
= \kappa(J) \frac{W_{\rm T}}{W}
,
\end{eqnarray}
where $\Delta\sigma_{\rm PT}$ is the P-odd T-odd cross section, $\Delta\sigma_{\rm P}$ the P-odd cross section, $W_{\rm T}$ the P-odd T-odd matrix element and $W$ the P-odd matrix element~\cite{gud17}. The calculations of these matrix elements were performed in Ref~\cite{fla92} and ~\cite{fla95}.
The spin factor $\kappa(J)$ is defined as
\begin{eqnarray}
\kappa(J) = \left\{ \begin{array}{cc}
	(-1)^{2I} \left(1+\frac{1}{2}\sqrt{\frac{2I-1}{I+1}}\frac{y}{x}\right) &\quad (J=I-\frac12) \\
	(-1)^{2I+1}\frac{I}{I+1} \left(1-\frac{1}{2}\sqrt{\frac{2I+3}{I}}\frac{y}{x}\right) &\quad (J=I+\frac12)
	\end{array}\right..
	\nonumber\\
\end{eqnarray}
The magnitude of $\kappa(J)$ indicates the sensitivity to the P-odd T-odd interaction.
The $J=I+\frac12$ case corresponds to the p-wave resonance of the ${}^{139}$La+n at $E_{\rm n}=E_2$.
The value of $\kappa(J)$ corresponding to the $\phi_2$ obtained is
\begin{eqnarray}
\kappa(J) = 4.84_{-1.69}^{+5.58},\quad 0.99_{-0.07}^{+0.08}
\end{eqnarray}
and $\abs{\kappa(J)}$ is shown in Fig.~\ref{fig:kappa}.


In the previous section, the $a_3$ term was ignored, as the centrifugal potential of the p-wave resonance is small. Hereafter we discuss the case when the $a_3$ term in Eq.~\ref{eq:Flambaum1} is activated. We analyze the angular dependences of $N_{\rm L}-N_{\rm H}$ and $N_{\rm L}+N_{\rm H}$ fitted by the functions of $f(\bar{P}_{d,1}/\bar{P}_{d,0})=A' \bar{P}_{d,1}/\bar{P}_{d,0}   + B'$ and $g(\bar{P}_{d,2}/\bar{P}_{d,0} )=C'\bar{P}_{d,2}/\bar{P}_{d,0}+D'$, respectively, with fitting parameters $A'$, $B'$, $C'$, and $D'$. The equations of $a_3$ can be written as  
\begin{figure}[t]
	\begin{center}
	\includegraphics[width=1.\linewidth]{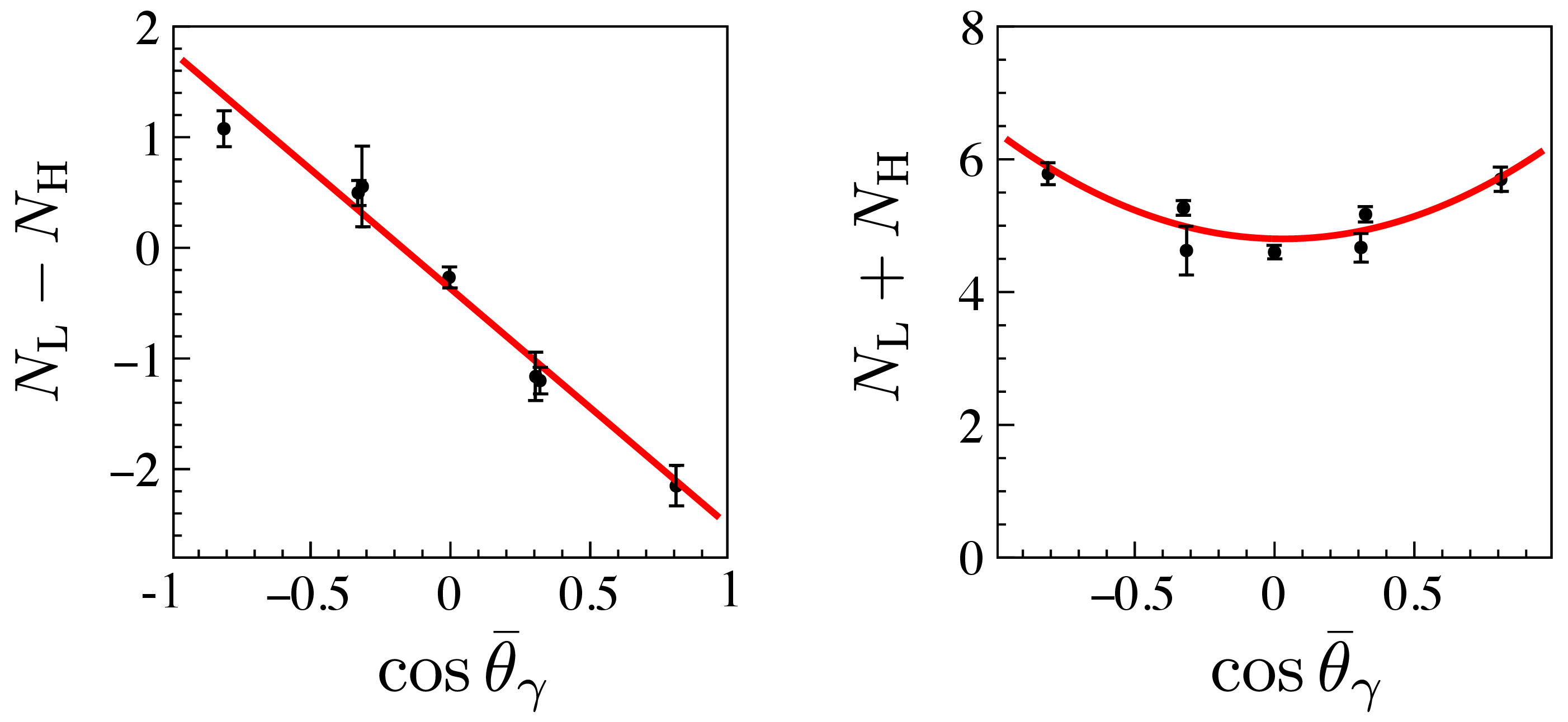}

	\caption[]{
	Angular dependences of $N_{\rm L}-N_{\rm H}$ and $N_{\rm L}+N_{\rm H}$. The solid line indicates the best fit. 
	}
	\label{nLnH}
	\end{center}
\end{figure}
\begin{eqnarray}
\frac{A'}{D'}&=&\frac{(\overline{a_1})_{\rm L}-(\overline{a_1})_{\rm H}}{(\overline{a_0})_{\rm L}+(\overline{a_0})_{\rm H}}\nonumber \\
&=&0.295 \cos\phi_2-0.345 \sin\phi_2,
\label{eq:Igamma2}
\\
\frac{C'}{D'}&=&\frac{(\overline{a_3})_{\rm L}+(\overline{a_3})_{\rm H}}{(\overline{a_0})_{\rm L}+(\overline{a_0})_{\rm H}}\nonumber \\
&=& -0.295\cos\phi_2\sin\phi_2 +0.050\sin^2\phi_2.
\label{eq:Igamma3}
\end{eqnarray}

The fitted results of $C'/D'$ and $A'/D'$ are
\begin{eqnarray}
\frac{C'}{D'}=0.191 \pm 0.028,\ \ \frac{A'}{D'}=-0.409 \pm  0.024.
\end{eqnarray}

\begin{figure}[hbtp]
	\centering
	\includegraphics[width=0.9\linewidth]{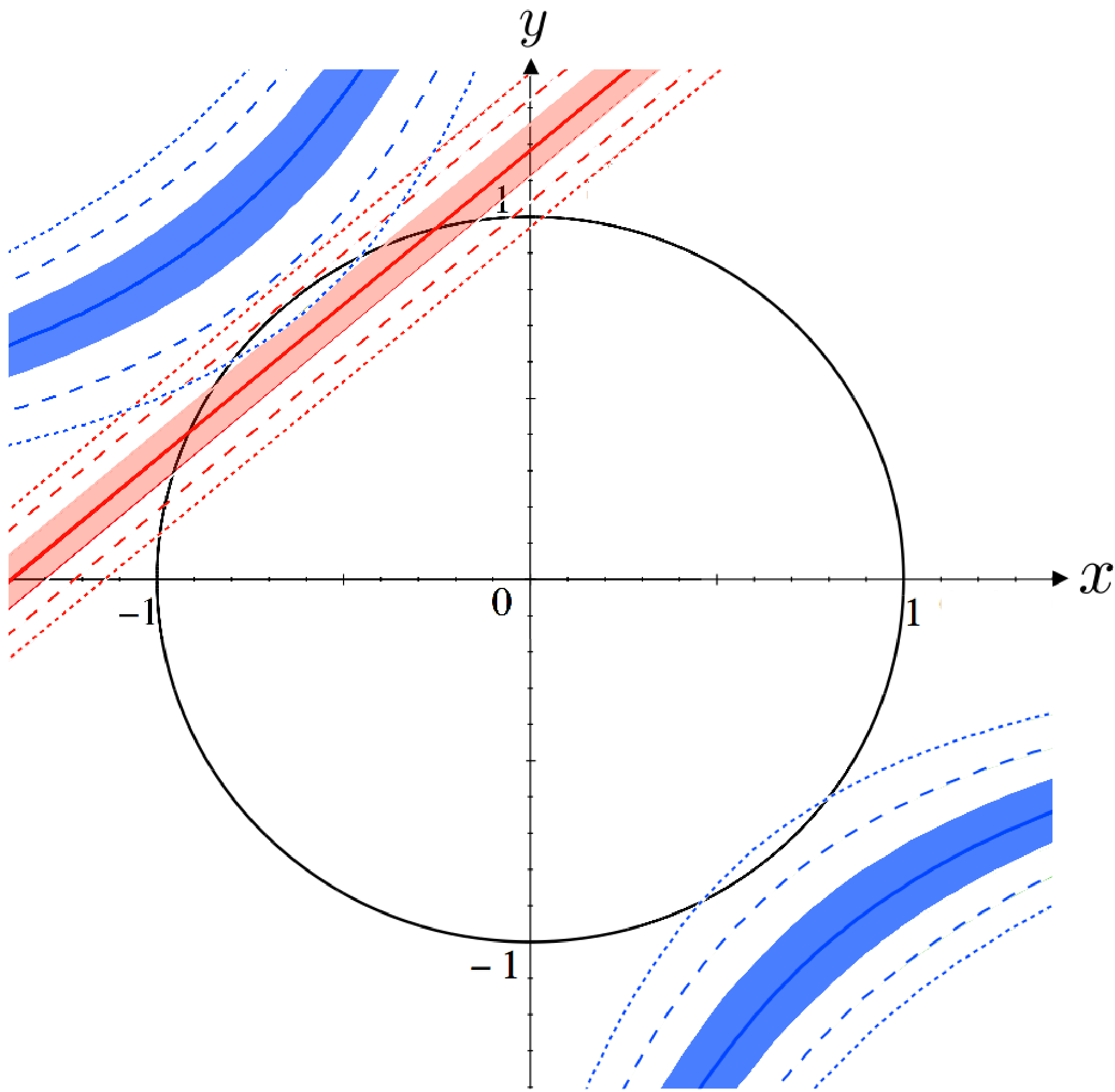}
	\caption[]{
	$a_1$(straight lines) and $a_3$(curved lines) on the $xy$-plane for the cases of $J_1=J_2=4, J_3=3$. The solid line, shaded area, dashed line, and dotted line show the central values of $\phi_2$, 1$\sigma$ areas, 2$\sigma$ contours and 3$\sigma$ contours, respectively. 
	}
	\label{fig:a0a1a3}
\end{figure}

The value of $\phi_2$ is determined by combining the equation of $a_1$(Eq.~\ref{eq:Igamma2}) and $a_3$(Eq.~\ref{eq:Igamma3}) on the $xy$-plane.
The result is shown in Fig.~\ref{fig:a0a1a3}.
The restriction from the $a_3$ term is not consistent with that of the $a_1$ term.
The $a_3$ term deviats from the requirement of $x_2^2+y_2^2=1$ by more than 2$\sigma$.\par
\begin{figure}[hbtp]
	\centering
	\includegraphics[width=0.9\linewidth]{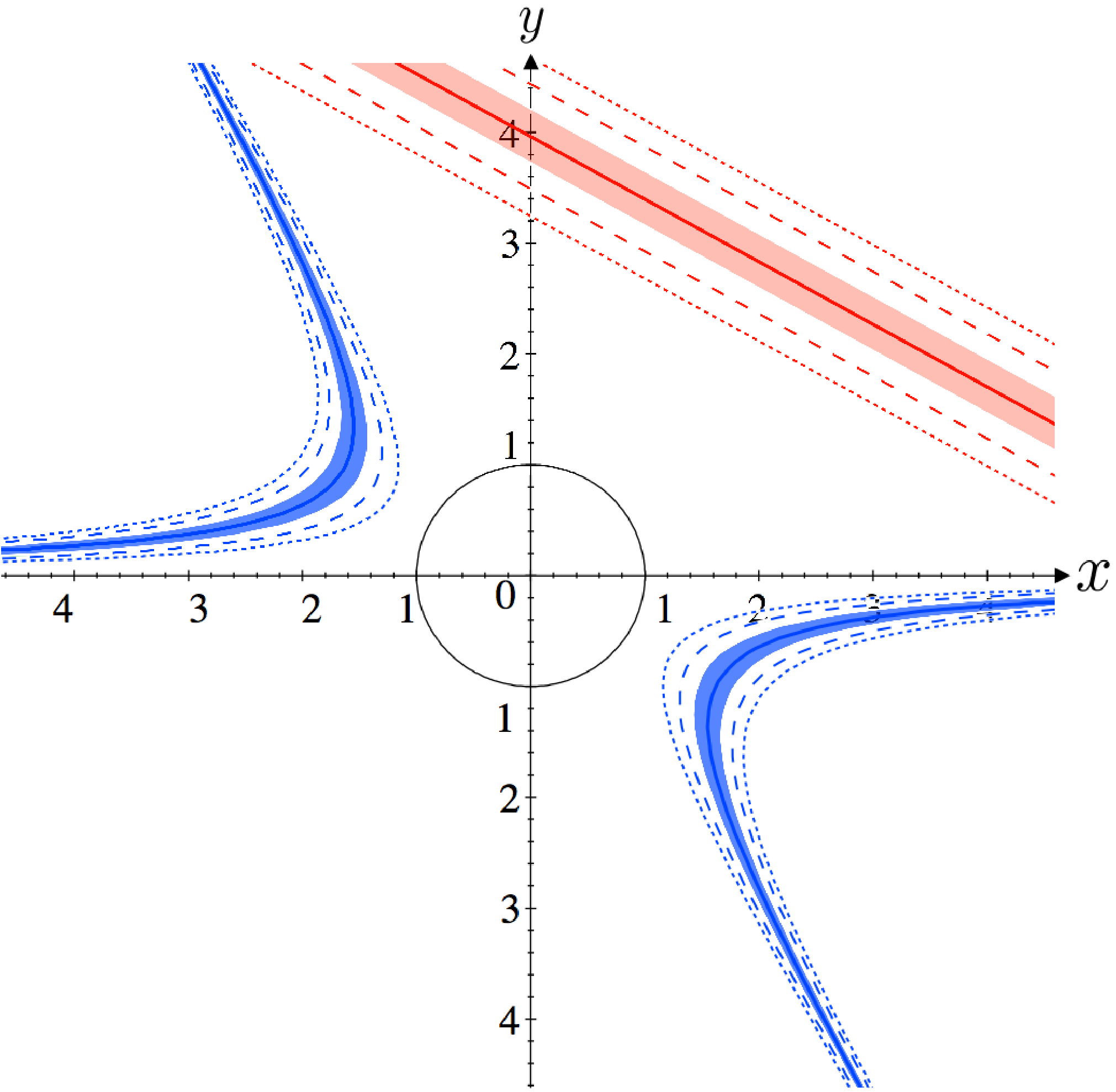}
	\caption[]{
	$a_1$(straight lines) and $a_3$(curved lines) on the $xy$-plane in the case of $J_1=J_2=J_3=3$. The solid line, shaded area, dashed line, and dotted line show the central values of $\phi_2$, 1$\sigma$ areas, 2$\sigma$ contours, and 3$\sigma$ contours, respectively.
	}
	\label{fig:a0a1a3_333}
\end{figure}

In this analysis, $J_1 = J_2 = 4, J_3 = 3$ are assumed. However, there is a possibility of the case of $J_1=J_2=J_3=3$. As the effect of the s$_2$-wave is negligibly small in this discussion, we discuss combinations of $J_1$ and $J_2$ only. The result of the case of $J_1=J_2=J_3=3$ is shown in Fig.~\ref{fig:a0a1a3_333}. 
Both $a_1$ and $a_3$ in the case of $J_1=J_2=J_3=3$ have no solution. As both $a_1$ and $a_3$ in the case of $J_1=J_2=4, J_3 = 3$ have solution in 3$\sigma$, we support $J_1=J_2=4$.\\
The origin of the inconsistency has not been identified in the present study. The inconsistency may be due to possible incompleteness of the reaction mechanism based on the interference between s- and p-wave amplitudes with the Breit--Wigner approximation.

\section{Conclusion}
We observed clear angular distribution of emitted $\gamma$-rays in the transition from the p-wave resonance of $^{139}$La+n to the ground state of $^{140}$La as a function of incident neutron energy. The angular distribution was analyzed by assuming interference between s- and p-wave amplitudes, and the partial neutron width of the p-wave resonance was obtained.
This result suggests that the T-violating effect can be enhanced on the same order of the P-violating effect for 0.74~eV p-wave resonance of $^{139}$La+n. Therefore an experiment to explore T-violation in compound nuclear states is feasible. 
In addition, the analysis under this assumption leads to results that are consistent with theoretical expectation, and we therefore believe the assumption of s-p mixing is correct.

\begin{acknowledgments}
The authors would like to thank the staff of ANNRI for the maintenance of the germanium detectors, and MLF and J-PARC for operating the accelerators and the neutron production target. We would like to thank Dr. K. Kino for the calculation of the pulse shape of neutron beam.
We also appreciate the continuous help by Prof. V.~P.~Gudkov for the interpretation of the measured results. The neutron scattering experiment was approved by the Neutron Scattering Program Advisory Committee of IMSS and KEK (Proposal No. 2014S03, 2015S12). The neutron experiment at the Materials and Life Science Experimental Facility of the J-PARC was performed under a user program (Proposal No. 2016B0200, 2016B0202, 2017A0158, 2017A0170, 2017A0203). This work was supported by MEXT KAKENHI Grant number JP19GS0210 and JSPS KAKENHI Grant Number JP17H02889.

\end{acknowledgments}

\begin{appendix}
\section{Definitions of symbols describing detector characteristics and the results of the simulation}
\begin{figure}[htbp]
	\centering
	\includegraphics[width=0.9\linewidth]{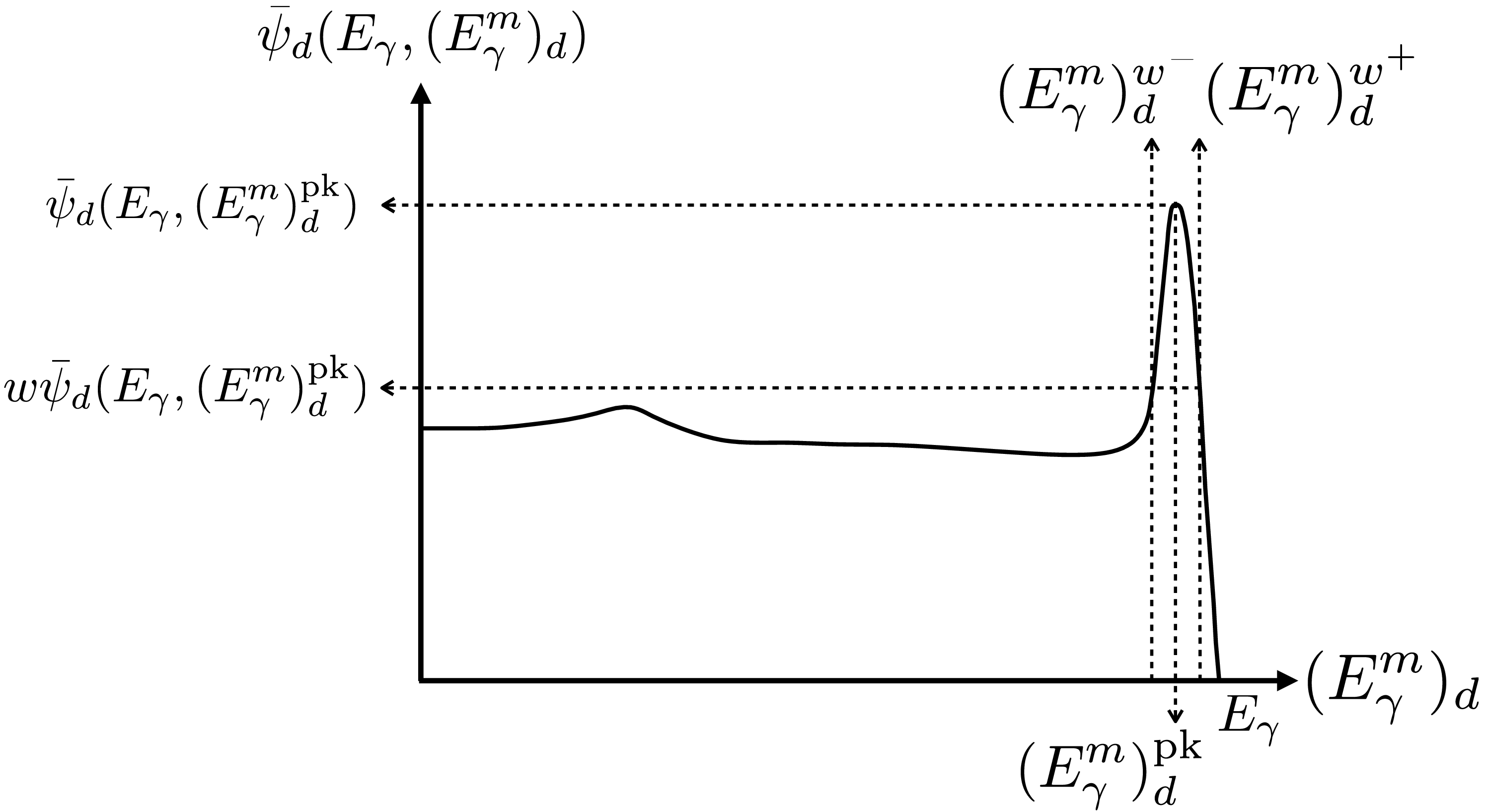}
	\caption[]{
	Schematic of the definition of the photo-peak in the pulse height spectrum.
	}
	\label{fig:PHSmodel}
\end{figure}

\begin{figure}[htbp]
	\centering
	\includegraphics[width=1.05\linewidth]{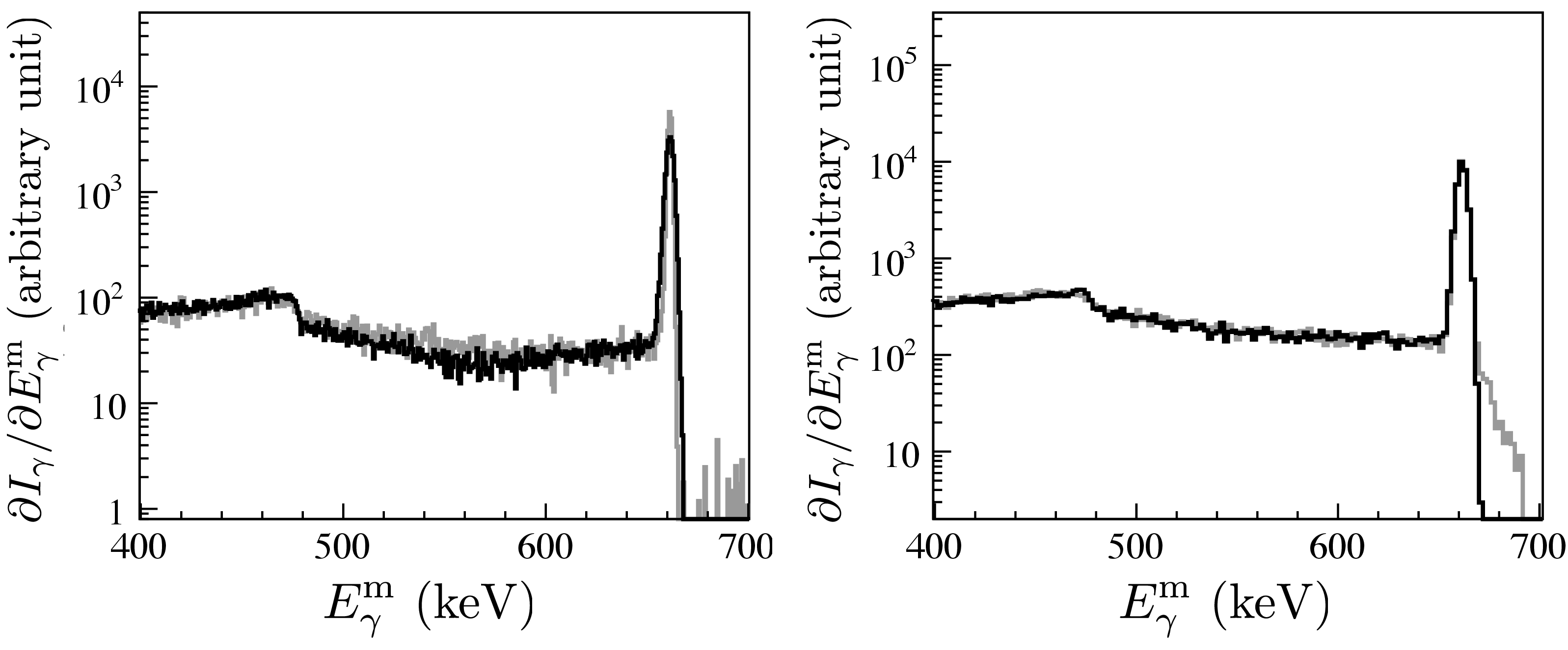}
	\caption[]{
	Comparison of the pulse height spectrum for $\gamma$-rays from a radioactive source ${}^{137}$Cs (gray line) and the numerical simulation  (black line) for a type-B detector unit (left) and a type-A detector unit (right). As the simulation reproduces the experimental data faithfully, these almost overlap. 
	}
	\label{Ge-Cs}
\end{figure}

\begin{figure}[htbp]
	\centering
	\includegraphics[width=1.05\linewidth]{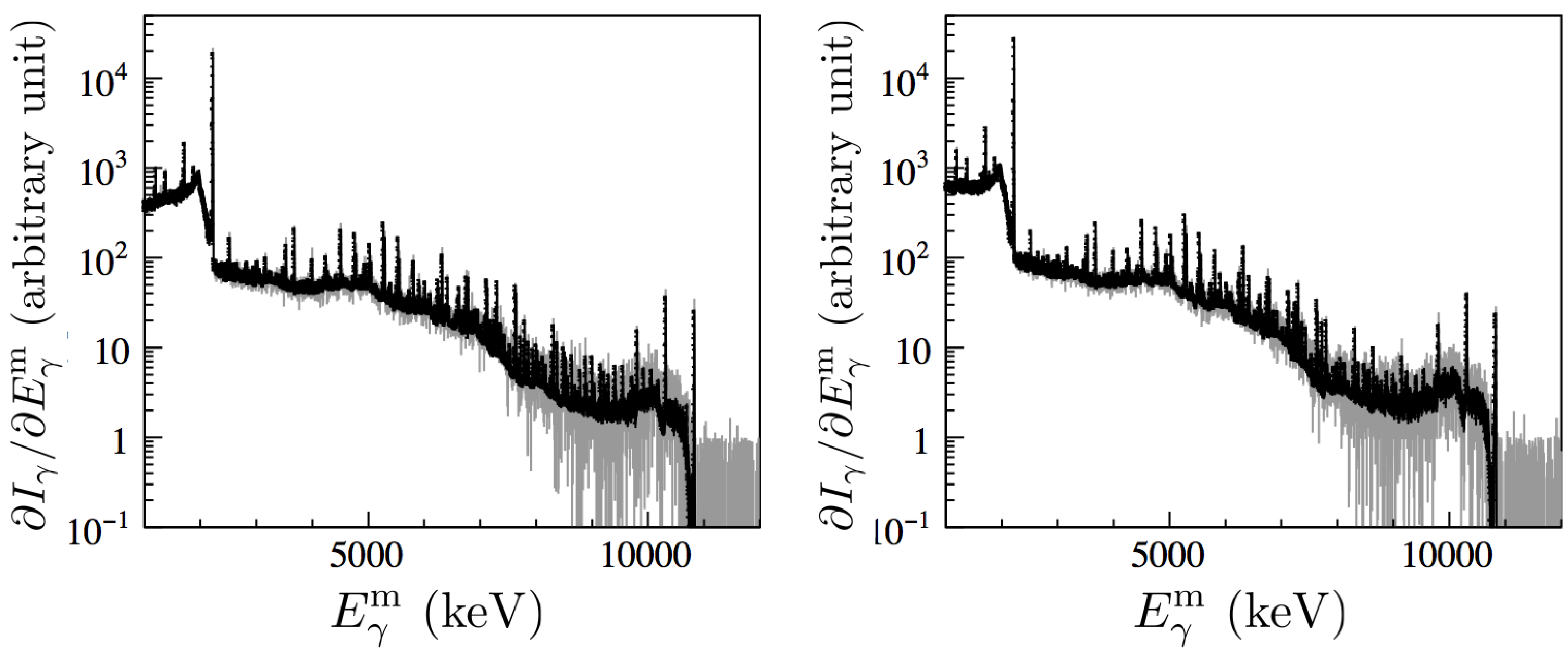}
	\caption[]{
	Comparison of pulse height spectrum for prompt $\gamma$-rays emitted in ${}^{14}$N(n,$\gamma$) reaction (gray line)  and the numerical simulation (black line) for a type-A detector unit (left) and a type-B detector unit (right). As the simulation reproduces the experimental data faithfully, these almost overlap. 
	}
	\label{Ge-Al}
\end{figure}

In this section, we describe the definition of characteristics of the germanium detectors and the simulation results.
Herein, we use $\psi_d(E_{\gamma},\Omega_{\gamma},(E^{\rm m}_{\gamma})_d)$ to denote the probability of the case where the energy of $(E^{\rm m}_{\gamma})_d$ is deposited in the $d$-th detector, when a $\gamma$-ray with an energy of $E_{\gamma}$ is emitted in the direction of $\Omega_{\gamma}$$=$$(\theta_{\gamma},\varphi_{\gamma})$. The polar angle and the azimuthal angle of the direction of the emitted $\gamma$-ray are denoted by  $\theta_\gamma$ and $\varphi_{\gamma}$, respectively.
The $\psi_d(E_{\gamma},\Omega_{\gamma},(E^{\rm m}_{\gamma})_d)$ satisfies
\begin{eqnarray}
\int_0^{E_{\gamma}} \psi_d(E_\gamma,\Omega_\gamma,(E^{\rm m}_{\gamma})_d) {\rm d}(E^{\rm m}_{\gamma})_d = 1
.
\end{eqnarray}
We define the distribution of the energy deposit as
\begin{eqnarray}
\bar{\psi}_d(E_{\gamma},(E^{\rm m}_{\gamma})_d)
=
\int_{\Omega_d}
	\psi_d(E_{\gamma},\Omega_{\gamma},(E^{\rm m}_{\gamma})_d)
	{\rm d}\Omega_{\gamma}
,
\label{eq:psibar}
\end{eqnarray}
where $\Omega_d$ is the geometric solid angle of the $d$-th detector.
The photo-peak efficiency of $d$-th detector for $\gamma$-rays with the energy of $E_\gamma$ is defined as
\begin{eqnarray}
\epsilon_d^{{\rm pk},w}(E_\gamma)
= \int_{(E^{\rm m}_{\gamma})_d^{w^{-}}}^{(E^{\rm m}_{\gamma})_d^{w^{+}}} \bar{\psi}_d(E_\gamma,(E^{\rm m}_{\gamma})_d) {\rm d}(E^{\rm m}_{\gamma})_d
,
\end{eqnarray}
where $(E^{\rm m}_{\gamma})_d^{w^{+}}$ and $(E^{\rm m}_{\gamma})_d^{w^{-}}$ are the upper and lower limits of the region of the energy deposit for defining the photo-peak region, as schematically shown in Fig.~\ref{fig:PHSmodel}.
For the definition of the photo-peak efficiency $w=1/4$ was used.
The relative photo-peak efficiency is also defined as 
\begin{eqnarray}
\bar{\epsilon}_d^{{\ \rm pk},w}(E_\gamma)
= \frac{\epsilon_d^{{\rm pk},w}(E_\gamma)}{\epsilon_1^{{\rm pk},w}(E_\gamma)}
.
\end{eqnarray}
The value of $\bar{\psi}_d(E_\gamma,(E^{\rm m}_{\gamma})_d)$ was obtained using the simulation to reproduce the pulse height spectra for $\gamma$-rays from the radioactive source of ${}^{137}$Cs ($E_{\gamma}$$=$$0.662\,{\rm MeV}$), as shown in Fig.~\ref{Ge-Cs}.
Subsequently, the reproducibility was checked by comparing the pulse height spectra for ${}^{60}$Co at $E_{\gamma}$$=$$1.173\,{\rm MeV}, 1.332\,{\rm MeV}$ and ${}^{22}$Na at $E_{\gamma}$=$1.275\,{\rm MeV}$.
Finally, we confirmed that the simulation program is applicable to higher energies by comparing the numerical simulation with the pulse height spectrum for prompt $\gamma$-rays emitted by the ${}^{14}$N(n,$\gamma$) reaction at $E_{\gamma}$$=$$10.829\,{\rm MeV}$, as shown in Fig.~\ref{Ge-Al}.
The photo-peak efficiency of the detector assembly for 1.332~MeV $\gamma$-ray is determined to be 3.64$\pm$0.11\%~\cite{kim09}.
In the case of angular distribution of $\gamma$-rays this is expanded using Legendre polynomials as $\sum_{p=0}^{\infty} c_p P_p(\cos\theta_\gamma)$, and the photo-peak counts of the $d$-th detector can be written as
\begin{eqnarray}
N_d(E_\gamma)&=&N_0 \sum_{p=0}^{\infty} c_p \bar{P}_{d,p}
,
\nonumber\\
\bar{P}_{d, p}&=&\frac{1}{4\pi}
\int_{(E^{\rm m}_{\gamma})_d^{w^{-}}}^{(E^{\rm m}_{\gamma})_d^{w^{+}}} {\rm d}(E^{\rm m}_{\gamma})_d
\int {\rm d}\Omega_\gamma
P_{p}(\cos\theta_\gamma)\bar{\psi}_d(E_\gamma,\Omega_\gamma)
.
\nonumber\\
\end{eqnarray}
The determiend values of $\bar{P}_{d,p}$ are listed in Table~\ref{tab:DetectorP} for $p=0,1,2$.
The quantity $\bar{P}_0$ corresponds to $\epsilon_d^{{\rm pk},w}$.
The table also contains the weighted average of the viewing angle of each detector $\bar\theta_d$ determined as 
$P_p(\cos\bar\theta_d)=\bar{P}_{d,p}/\bar{P}_{d,0}$.
\begin{table*}[htbp]
\begin{center}
	\small
	\scalebox{0.9}[0.9]{
	\begin{tabular}{c||c|c||c||c|c|c||c}
	\multirow{2}{*}{$d$} &
	\multicolumn{3}{c||}{$E_{\gamma}=0.662\,{\rm MeV}$} &
	\multicolumn{4}{c}{$E_{\gamma}=5.262\,{\rm MeV}$}
	\\
	\cline{2-8}
	&  $\bar{P}_{d,1}/\bar{P}_{d,0}$ &  $\bar{P}_{d,2}/\bar{P}_{d,0}$ & $\bar{\theta}_d$
	& $\bar{\epsilon}_d^{{\rm pk},1/4}$ & $\bar{P}_{d,1}/\bar{P}_{d,0}$ & $\bar{P}_{d,2}/\bar{P}_{d,0}$ & $\bar{\theta}_d$
	\\
	\hline
	
	$1$  & $0.00001\pm0.00002$ & $-0.488480\pm0.000003$ & $90.000\pm0.001$ & $1$ & $-0.00004\pm0.00004$ & $-0.49036\pm0.00001$ & $90.002\pm0.002$ \\
	$2$  & $-0.00003\pm0.00003$ & $-0.489485\pm0.000003$ & $90.002\pm0.001$ & $0.997\pm0.027$ &$0.00006\pm0.00004$ & $-0.49115\pm0.00001$ & $89.997\pm0.002$ \\
	$3$  & $0.30297\pm0.00002$ & $-0.35283\pm0.00002$ & $72.299\pm0.001$ & $0.920\pm0.026$ & $0.29887\pm0.00004$ & $-0.35791\pm0.00003$ & $72.556\pm0.002$ \\
	$4$  & $0.30294\pm0.00002$ & $-0.35285\pm0.00002$ & $72.301\pm0.001$ & $1.041\pm0.028$ & $0.29889\pm0.00004$ & $-0.35791\pm0.00003$ & $72.555\pm0.002$\\
	$5$  & $0.00008\pm0.00002$ & $-0.489493\pm0.000003$ & $89.996\pm0.001$ & $1.025\pm0.028$ & $-0.00012\pm0.00004$ & $-0.49116\pm0.00001$ & $90.007\pm0.002$ \\
	$6$  & $-0.30293\pm0.00002$ & $-0.35288\pm0.00002$ & $107.698\pm0.001$ & $1.092\pm0.029$ & $-0.29886\pm0.00004$ & $-0.35792\pm0.00003$ & $107.443\pm0.002$ \\
	$7$  & $-0.30297\pm0.00002$ & $-0.35284\pm0.00002$ & $107.700\pm0.001$ & $0.999\pm0.027$ & $-0.29885\pm0.00004$ & $-0.35794\pm0.00003$ & $107.442\pm0.002$\\
	$8$  & $0.00004\pm0.00002$ & $-0.488259\pm0.000003$ & $89.998\pm0.001$ & $0.923\pm0.026$ & $0.00005\pm0.00004$ & $-0.49030\pm0.00001$ &$89.997\pm0.002$ \\
	$9$  & $-0.00002\pm0.00003$ & $-0.48927\pm0.000003$ & $90.001\pm0.001$ & $0.915\pm0.026$ & $-0.00003\pm0.00004$ & $-0.49107\pm0.00001$ & $90.002\pm0.002$ \\
	$10$  & $0.30292\pm0.00002$ & $-0.35288\pm0.00002$ & $72.302\pm0.001$ & $0.998\pm0.027$ &$0.29894\pm0.00004$ & $-0.35785\pm0.00003$ & $72.552\pm0.002$ \\
	$11$  & $0.30297\pm0.00002$ & $-0.35284\pm0.00002$ & $72.300\pm0.001$ & $0.945\pm0.026$ & $0.29886\pm0.00004$ & $-0.35793\pm0.00003$ & $72.557\pm0.002$ \\
	$12$  & $-0.00001\pm0.00003$ & $-0.489256\pm0.000003$ & $90.001\pm0.001$ & $0.979\pm0.027$ & $0.00004\pm0.00004$ & $-0.49108\pm0.00001$ & $89.997\pm0.002$ \\
	$13$  & $-0.30293\pm0.00002$ & $-0.35286\pm0.00002$ & $107.698\pm0.001$ & $1.046\pm0.028$ & $-0.29892\pm0.00004$& $-0.357867\pm0.00003$ & $107.447\pm0.002$ \\
	$14$  & $-0.30297\pm0.00002$ & $-0.35284\pm0.00002$ & $107.700\pm0.001$ & $1.142\pm0.030$ & $-0.29887\pm0.00004$ & $-0.35791\pm0.00003$ & $107.444\pm0.002$ \\
	$15$  & $-0.80431\pm0.00001$ & $0.47410\pm0.00004$ & $143.824\pm0.001$ & $0.892\pm0.025$ & $-0.80420\pm0.00002$ & $0.47340\pm0.00005$ & $143.780\pm0.002$ \\
	$16$  & $-0.30772\pm0.00003$ & $-0.35142\pm0.00003$& $107.967\pm0.002$ & $0.356\pm0.014$ & $-0.30775\pm0.00004$ & $-0.35229\pm0.00004$ & $107.963\pm0.003$ \\
	$17$  & $0.30697\pm0.00002$ & $-0.34904\pm0.00002$ & $72.058\pm0.001$ & $0.945\pm0.026$ & $0.30715\pm0.00004$ & $-0.34994\pm0.00003$ & $72.054\pm0.002$ \\
	$18$  & $0.80434\pm0.00001$ & $0.47419\pm0.00004$ & $36.173\pm0.001$ & $1.170\pm0.031$ & $0.80418\pm0.00002$ & $0.47337\pm0.00005$ & $36.221\pm0.002$ \\
	$19$  & $0.80425\pm0.00001$ & $0.47396\pm0.00004$ & $36.182\pm0.001$ & $0.917\pm0.026$ & $0.80415\pm0.00002$ & $0.47328\pm0.00005$ & $36.225\pm0.002$ \\
	$20$  & $0.30683\pm0.00002$ & $-0.34917\pm0.00002$& $72.066\pm0.001$ & $$ & $0.30705\pm0.00004$ & $-0.35004\pm0.00003$ & $72.060\pm0.002$ \\
	$21$  & $-0.30689\pm0.00002$ & $-0.34911\pm0.00002$ & $107.938\pm0.001$ & $$ & $-0.30709\pm0.00004$ & $-0.34999\pm0.00003$ & $107.943\pm0.002$ \\
	$22$  & $-0.80427\pm0.00001$ & $0.47401\pm0.00004$ & $143.820\pm0.001$ & $1.113\pm0.030$ & $-0.80420\pm0.00002$ & $0.47341\pm0.00005$ & $143.781\pm0.002$ \\	
	\end{tabular}
	}
	\caption{
	Values of $\bar{P}_{d,p}/\bar{P}_{d,0}$ for $p=1,2$ are shown for all detectors together with the weighted average angle of each detector $\bar\theta_d$ at $E_{\gamma}$=0.662\ MeV and $E_{\gamma}$=5.262\ MeV. The  relative peak efficiency : $\bar{\epsilon}_d^{{\rm pk},1/4}$ was measured using $^{14}$N(n,$\gamma$) reactions are also shown.
	}
	\label{tab:DetectorP}
\end{center}	
\end{table*}

\section{Neutron absorption cross section}
The formula used to describe the neutron cross section $\sigma_{\rm t}$ is given as a function of the neutron energy in the center-of-mass system:
\begin{eqnarray}
\sigma_{\rm t}(E) = \sigma_{\rm s} + \sigma_{\rm n\gamma}(E_{\rm{n}})
,
\end{eqnarray}
\begin{eqnarray}
\sigma_{\rm s} = 4\pi a^2
,
\end{eqnarray}
\begin{eqnarray}
\Gamma_r^{\gamma}
	&=& \sum_f \Gamma_{r,f}^{\gamma}
	,
	\\
\sigma_{\rm n\gamma}(E_{\rm n})
	&=& \sum_f \sigma_{{\rm n}\gamma_f}(E_{\rm n})
	,
	\\
\sigma_{{\rm n}\gamma_f}(E_{\rm n})
	&=& \sum_r \frac{\pi\hbar^2}{2m_{\rm n}E_{\rm n}} \left[\frac{E_{\rm n}}{E_r}\right]^{l_r+1/2}
	\nonumber\\&&\quad\times
	\frac
		{\displaystyle g_r \Gamma_r^{\rm n} \Gamma_{r,f}^{\gamma}}
		{\displaystyle (E_{\rm n}-E_r)^2+(\Gamma_r/2)^2}
	,
\end{eqnarray}
where $\sigma_{\rm s}$ is the scattering cross section, $\sigma_{\rm n\gamma}$ is the radiative capture cross section, $k_{\rm n}=\sqrt{2m_{\rm n}E_{\rm n}}/\hbar$, $a$ are scattering length, $E_r$ is the resonance energy of the $r$-th resonance, $g_r=(2J_r+1)/(2(2I+1))$ is the statistical factor with the target nuclear spin $I$ and the spin of the $r$-th resonance $J_r$, $\Gamma_r^{\rm n}$ is the neutron width of $r$-th resonance, and $\Gamma_{r,f}^{\gamma}$ is the $\gamma$-width of the $\gamma$-ray transition from the $r$-th resonance to the final state $f$.
The quantity $l_r$ represents the orbital angular momentum of the incident neutrons contributing to $r$-th resonance.
The above formula is valid for $l_r=0, 1$.
Higher angular momenta can be ignored in the incident neutron energy region of our interest.

\section{Pulse shape of the neutron beam}
The pulse shape of neutron beam depends on the neutron energy $E_{\rm n}$ according to the time delay during the moderation process.
The double differential of the flux of the pulsed neutron beam $I_{\rm{n}}$ as a function of $E_{\rm n}$, and the time measured from the primary proton beam injection $t$ is known to be well reproduced by the Ikeda--Carpenter function, defined as
\begin{eqnarray}
&&\frac{\partial^2 I_{\rm{n}}}{\partial E_{\rm n} \partial t}
(E_{\rm n},t)
=
\frac{\alpha C}{2} \left\{
	(1-R) (\alpha t)^2 e^{-\alpha t} + 2R\frac{\alpha^2\beta}{(\alpha-\beta)^3} \right.
	\nonumber\\ && \quad\quad\quad \times \left.
	\left[ e^{-\beta t}-e^{-\alpha t}\left( 1 + (\alpha-\beta)t+\frac{(\alpha-\beta)^2}{2}t^2 \right) \right]
	\right\}
	,
	\nonumber\\
\label{eq:IkedaCarpenter}
\end{eqnarray}
where parameters $\alpha$, $\beta$, $C$, $R$ depend on $E_{\rm n}$.
The Ikeda--Carpenter function was originally proposed to explain the pulse shape of the cold source of polyethylene moderator at the Intense Pulsed Neutron Source of the Argonne National Laboratory~\cite{ike85}.
The double differential of the neutron beam flux on the moderator surface of the J-PARC spallation source was calculated using MCNPX~\cite{JSNS}, and was fitted with the Ikeda--Carpenter function. The dependence of neutron energy on the fitting parameters $t_0, \alpha, \beta, R$, and C were obtained by fitting of the pulse shape of neutron beam with a polynomial function\cite{kin11}.

The energy spectrum at a given time $t^{\rm m}$ at the distance of $L$ from the moderator surface is given as
\begin{eqnarray}
\pdiff{I_{\rm{n}}}{t^{\rm m}} (t^{\rm m})
&=&
\int {\rm d}E'
	\frac{\partial^2 I_{{\mathrm n}}}{\partial E_{\rm n} \partial t}
	\left(E',t^{\rm m} - L\sqrt{\frac{m_{\rm n}}{2E'}}\right)
	.
	\nonumber\\
\end{eqnarray}

\section{Thermal motion of the target nucleus}
We adopted the free gas model for the thermal motion of the target nuclei, which leads to the $\gamma$-ray yield in the form of
\begin{eqnarray}
\pdiff{I_{\gamma}}{t^{\rm m}} (t^{\rm m})
&=&
I_0 \int
	{\rm d}E'
	{\rm d}^3p_{\rm A}
	\,
	\frac{\partial^2 I_{\rm{n}}}{\partial E_{\rm n} \partial t}
		\left(E',t^{\rm m} - L \sqrt{\frac{m_{\rm n}}{2E'}}\right)
	\nonumber\\ &&\times
	\frac{1}{(2\pi m_{\rm A}k_{\rm B}T)^{3/2}} e^{-p_{\rm A}^2/2m_{\rm A}k_{\rm B}T}
	\nonumber\\ &&\times
	\frac
		{\sigma_{\rm n\gamma}(E_{\rm n})}
		{\sigma_{\rm t}(E_{\rm n})}
	\left(1 - e^{-n\sigma_{\rm t}(E)\,\Delta z}\right)
	,
	\nonumber\\
\label{eq:Igamma}
\end{eqnarray}
as long as the target is sufficiently thin, such that multiple scattering is negligible.
$I_0$ is the normalization constant, $\Delta z$ is the target thickness, $n$ is the number density of target nuclei, $k_{\rm B}$ is the Boltzmann constant, and $T$ is the effective temperature of the target, which can be used as a fitting parameter.

\section{Formula describing the (n,$\gamma$) angular dependence}
The differential cross section of the (n,$\gamma$) reaction induced by unpolarized neutrons can be written as
\begin{eqnarray}
\diff{\sigma_{{\rm n}\gamma_f}}{\Omega_\gamma}
	&=& \frac{1}{2}\left(
	a_{0}+a_{1}\cos\theta_{\gamma}+a_{3}\left(\cos^2\!\!\theta_{\gamma}-\frac{1}{3}\right)
	\right)
	,
	\nonumber\\
a_0 &=& \sum_{r_{\rm s}} \abs{V_{r_{\rm s}}}^2 + \sum_{r_{\rm p}} \abs{V_{r_{\rm p}}}^2
	,\nonumber\\
a_1 &=& 2 {\rm{Re}} \sum_{r_{\rm s}i_{\rm p}j} V_{r_{\rm{s}}} V_{r_{\rm p}}^{\ast} z_{r_{\rm p}j} P(J_{r_{\rm s}}J_{r_{\rm p}}\frac{1}{2}j1IF)
	,\nonumber\\
a_3 &=& 3\sqrt{10}\ {\rm{Re}} \sum_{r_{\rm p} j r_{\rm p}^{\prime} j^{\prime}} V_{r_{\rm p}} V_{r_{\rm p}^{\prime}}^{\ast} z_{r_{\rm p}j} z_{r_{\rm p}j'}
	\nonumber\\&&
	\times P(J_{r_{\rm p}} J_{r_{\rm p}^{\prime}} j j^{\prime} 2 I F) 
	\nineJ{2}{1}{1}{0}{1/2}{1/2}{2}{r_{\rm p}}{j'}
	\nonumber\\
	\label{eq:Flambaum1}
\end{eqnarray}
\begin{eqnarray}
&&
V_{r_{\rm s}f} = -\frac{\sqrt{g_{r_{\rm s}}}}{2k_{\rm n}}\frac{\sqrt{\Gamma_{r_{\rm s}}^{\rm n} \Gamma_{r_{\rm s}f}^{\gamma}} (1+\alpha_{r_{\rm s}})}{E_{\rm n}-E_{r_{\rm s}}+i\Gamma_{r_{\rm s}}/2}
	\nonumber\\
&&
V_{r_{\rm p}f} = -\frac{\sqrt{g_{r_{\rm p}}}}{2k_{\rm n}}
	\frac{\sqrt{\Gamma_{r_{\rm p}}^{\rm n} \Gamma_{r_{\rm p}f}^{\gamma}}}
	{E_{\rm n}-E_{r_{\rm p}}+i\Gamma_{r_{\rm p}}/2}
	\nonumber\\
&&
z_{r_{\rm p} j} = \sqrt{\frac{\Gamma_{r_{\rm p}j}^{\rm n}}{\Gamma_{r_{\rm p}}^{\rm n}}}
	= \left\{ \begin{array}{ll}
		x_{r_{\rm p}} & \quad (j=1/2) \\
		y_{r_{\rm p}} & \quad (j=3/2)
	\end{array} \right.
	,
	\nonumber\\
&&
P(JJ'jj'kIF) = (-1)^{J+J'+j'+I+F}
	\nonumber\\&&\quad\quad\quad\quad\quad\quad\quad\times
	\frac{3}{2}\sqrt{(2J+1)(2J'+1)(2j+1)(2j'+1)}
	\nonumber\\&&\quad\quad\quad\quad\quad\quad\quad\times
	\sixJ{k}{j}{j'}{I}{J'}{J}
	\sixJ{k}{1}{1}{F}{J}{J'}
	,
	\nonumber\\
	\label{eq::flambaum2}
\end{eqnarray}
where $r_{\rm s}$ is the r$^{\text{th}}$ s-wave resonance number ($l_{r_{\rm s}}$$=$$0$) and $r_{\rm p}$ is the r$^{\text{th}}$ p-wave resonance number($l_{r_{\rm p}}$$=$$1$).
Amplitudes $V_{r_{\rm s}}$ and $V_{r_{\rm p}}$ are the s- and p-wave amplitudes, respectively, and $\alpha_{r_{\rm s}}$ represents the contribution from far s-wave resonances.
Width $\Gamma_{r_{\rm p}j}^{\rm n}$ is the partial neutron width for the incident neutrons of total angular momentum of $j$, and $x_{r_{\rm p}}$ and $y_{r_{\rm p}}$ are defined as 
\begin{eqnarray}
x_{r_{\rm p}}= \sqrt{\frac{\Gamma_{\mathrm{p},j=1/2}^{\rm n}}{\Gamma^{\rm n}_{\rm p}}},\ y_{r_{\rm s}}= \sqrt{\frac{\Gamma_{\mathrm{ p},j=3/2}^{\rm n}}{\Gamma^{\rm n}_{\rm p}}}.
\end{eqnarray}
$x_{r_{\rm p}}$ and $y_{r_{\rm p}}$ satisfy 
\begin{eqnarray}
x_{r_{\rm p}}^2+y_{r_{\rm p}}^2=1
\end{eqnarray}
due to the relation $\Gamma_{\rm p}^{\rm n}=\Gamma_{{\rm p},j=1/2}^{\rm n}+\Gamma_{{\rm p},j=3/2}^{\rm n}$.
The resonance energy and the resonance width obtained from this experiment, the published values listed in Table~\ref{tab:Igamma-fit}, and $I=7/2$ are used to determine the value of $\phi_{r_{\rm p}}$, defined as
\begin{eqnarray}
x_{r_{\rm p}} = \cos\phi_{r_{\rm p}} , \quad y_{r_{\rm p}} = \sin\phi_{r_{\rm p}} .\label{eq:xandy}
\end{eqnarray}
In the case of $^{139}$La, negative s-wave amplitude $V_1$, p-wave amplitude $V_2$, and positive s-wave amplitude $V_3$ can be written as,
 \begin{eqnarray}
V_{1f} &=& - \lambda_{1f}
	\left(\frac{\abs{E_1}}{E_{\rm n}}\right)^{\frac14}
	\frac
		{\Gamma_1/2}
		{E_{\rm n}-E_1+i\Gamma_1/2}
	,\nonumber\\
V_{2f} &=& - \lambda_{2f}
	\left(\frac{E_{\rm n}}{E_2}\right)^{\frac14}
	\frac
		{\Gamma_2/2}
		{E_{\rm n}-E_2+i\Gamma_2/2}
	,\nonumber\\
V_{3f} &=& - \lambda_{3f}
	\left(\frac{E_3}{E_{\rm n}}\right)^{\frac14}
	\frac
		{\Gamma_3/2}
		{E_{\rm n}-E_3+i\Gamma_3/2}
	,
\nonumber\\
\end{eqnarray}
where the absolute value of $E_1$ is adopted simply to avoid the imaginary neutron width.
The terms $a_0$, $a_1$, and $a_3$ are given as
\begin{eqnarray}
a_0
	&=&
	\lambda_{1f}^2
		\sqrt{\frac{\abs{E_1}}{E_{\rm n}}}
		\frac{\Gamma^{\ 2}_1/4}{(E_{\rm n}-E_1)^2+\Gamma_1^{\ 2}/4}
	\nonumber\\&&
	+
	\lambda_{2f}^2
		\sqrt{\frac{E_{\rm n}}{E_2}}
		\frac{\Gamma_2^{\ 2}/4}{(E_{\rm n}-E_2)^2+\Gamma_2^{\ 2}/4}
	\nonumber\\&&
	+
	\lambda_{3f}^2
		\sqrt{\frac{E_3}{E_{\rm n}}}
		\frac{\Gamma_3^{\ 2}/4}{(E_{\rm n}-E_3)^2+\Gamma_3^{\ 2}/4}
	\nonumber\\
a_1
	&=&
		\lambda_{1f}\lambda_{2f}
		\left(\frac{\abs{E_1}}{E_{2}}\right)^{\frac{1}{4}}
		\nonumber\\&&\times
		\frac
			{\Gamma_1\Gamma_2(E_{\rm n}-E_1)(E_{\rm n}-E_2)+\Gamma_1^{\ 2}\Gamma_2^{\ 2}/4}
			{
				2
				\left((E_{\rm n}-E_1)^2+\Gamma_1^{\ 2}/4\right)
				\left((E_{\rm n}-E_2)^2+\Gamma_2^{\ 2}/4\right)
			}
		\nonumber\\&&\times
		\frac5 8
		\left(- x + \sqrt{\frac75} y \right)
	\nonumber\\&&
	+
		\lambda_{3f}\lambda_{2f}
		\left(\frac{E_3}{E_{2}}\right)^{\frac{1}{4}}
		\nonumber\\&&\times
		\frac
			{\Gamma_3\Gamma_2(E_{\rm n}-E_3)(E_{\rm n}-E_2)+\Gamma_3^{\ 2}\Gamma_2^{\ 2}/4}
			{
				2
				\left((E_{\rm n}-E_3)^2+\Gamma_3^{\ 2}/4\right)
				\left((E_{\rm n}-E_2)^2+\Gamma_2^{\ 2}/4\right)
			}
		\nonumber\\&&\times
		\frac{3\sqrt{3}}{8}
		\left( x + \sqrt{\frac{5}{7}}y\right)
	\nonumber\\
a_3
	&=&
	\lambda_{2f}^2
		\sqrt{\frac{E_{\rm n}}{E_2}}
		\frac{\Gamma_2^{\ 2}/4}{(E_{\rm n}-E_2)^2+\Gamma_2^{\ 2}/4}
		\frac{33}{280}
		\left(
			-\sqrt{35} xy
			+
			 y^2
		\right)
	.
	\nonumber\\
\end{eqnarray}
It can be assumed that the energy dependence of the neutron width of the $r$-th resonance is given as
\begin{eqnarray}
\Gamma_r^{\rm n} =
 (k_{\rm n}R)^{2l_r} \sqrt{\frac{E_{\rm n}}{1\,{\rm eV}}}
 \Gamma_r^{{\rm n}l_r}\label{reducedwidth}
\end{eqnarray}
for $E_{\rm n}>0$ where $R$ is the radius of target nuclei and $\Gamma_r^{{\rm n}l_r}$ is the reduced neutron width.
This energy dependence is implemented as
\begin{eqnarray}
\Gamma_r^{\rm n} = \left(\frac{E_{\rm n}}{\abs{E_r}}\right)^{l_r+\frac{1}{2}} \overline{\Gamma_r^{\rm n}}
,
\end{eqnarray}
where $\overline{\Gamma_{\rm r}^{\rm n}}$ is a constant independent of the energy .
As the phase shift due to the optical potential is negligibly small, each amplitude can be written as
\begin{eqnarray}
V_{rf}
	&=&
		- \lambda_{rf}
		\left(\frac{E_{\rm n}}{\abs{E_r}}\right)^{\frac{l_r}{2}-\frac{1}{4}}
		\frac{\Gamma_r/2}{E_{\rm n}-E_r+i\Gamma_r/2}
	\nonumber\\
	.
\end{eqnarray}
where $\lambda_{rf}$ is defined as
\begin{eqnarray}
\lambda_{rf} = 
	\frac{\hbar}{2}
	\sqrt{\frac
		{2g_r\overline{\Gamma_r^{\rm n}}\Gamma_{rf}^{\rm\gamma}}
		{m_{\rm n}\abs{E_r}\Gamma_r^2}
	}
	.
\end{eqnarray}
$\Gamma_{rf}^{\rm\gamma}$ is the $\gamma$-width from the $r$-th resonance to the final state.

\end{appendix}

\newpage	

\newpage	

\newpage	
\clearpage

\section*{Erratum}
%


We correct the values of the spin factor $\kappa(J)$ in our previous paper.  
The spin factor $\kappa(J)$ was originally given by Eq.~23 in Ref.~\cite{gud17} as a function of the nuclear spin and the neutron partial widths defined by $x$ and $y$. In the original paper, the values of $x$ and $y$ were obtained from the analysis result based on the formalism by Flambaum {\it{et al.}}~\cite{fla85}. However, due to the different order of summation of the neutron spin and neutron orbital angular momentum, the sign of $y$ defined by Gudkov {\it{et al.}}~\cite{gud17} was different to that defined by Flambaum {\it{et al.}}~\cite{fla85}. Therefore, $x$ and $y$ in Eq.~22 in the original paper should be replaced with $x \rightarrow x$ and $y \rightarrow -y$, and Eq.~22 in the original paper should be corrected as, 
\begin{align}
\kappa(J) = \left\{ \begin{array}{cc}
	\left(1-\frac{1}{2}\sqrt{\frac{2I-1}{I+1}}\frac{y}{x}\right) &\quad (J=I-\frac12) \\
	\frac{I}{I+1} \left(1+\frac{1}{2}\sqrt{\frac{2I+3}{I}}\frac{y}{x}\right) &\quad (J=I+\frac12)
	\end{array}\right..
	\nonumber\tag{1} 
\end{align}

\begin{figure}[hbtp]
	\centering
		\includegraphics[width=0.9\linewidth]{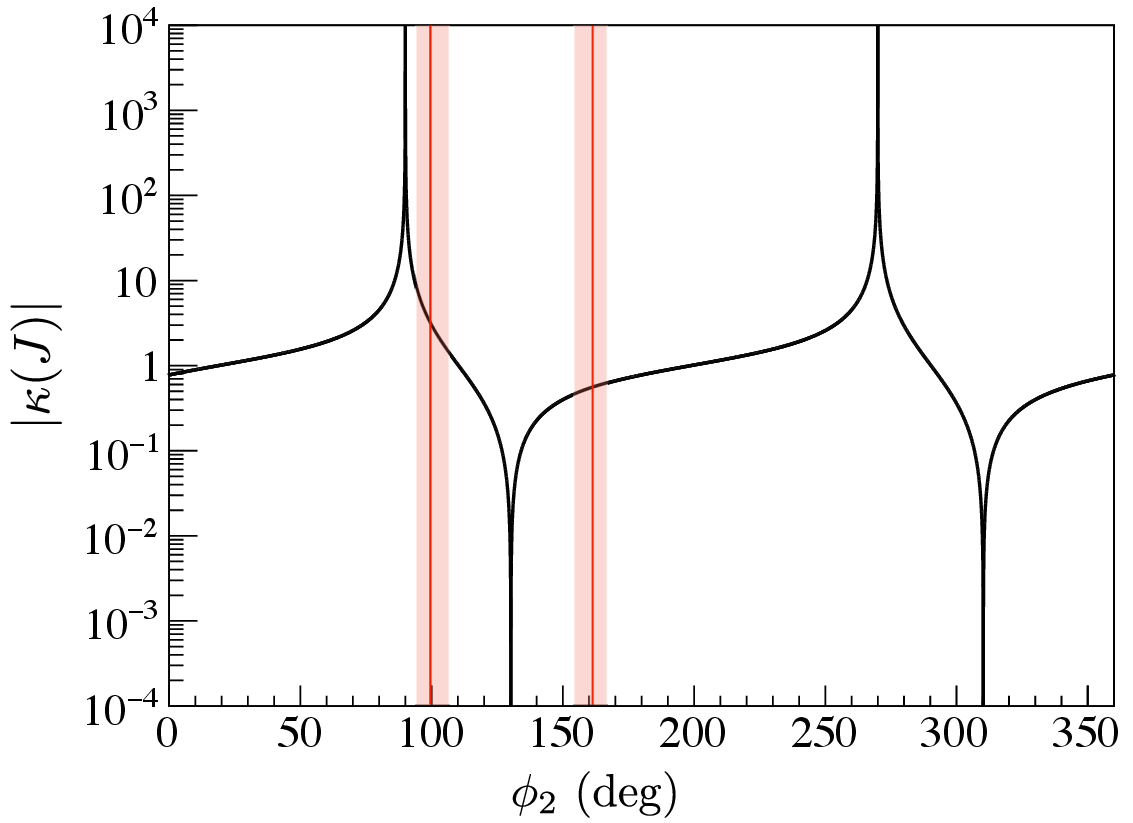}
	\caption[]{
	Value of $\abs{\kappa(J)}$ as a function of $\phi_2$. The solid line and shaded area show the central values of $\phi_2$ and the 1$\sigma$ area from central value, respectively. 
	}
	\label{fig:kappa_rev}
\end{figure}

Consequently, the $\kappa(J)$ values in Eq.~23 in original paper are corrected as 
\begin{align}
\kappa(J) = -3.28_{-5.58}^{+1.69},\quad 0.56_{-0.08}^{+0.07} \tag{2}
\end{align}
Similarly, Fig.~19 in the original paper should be replaced as Fig.~\ref{fig:kappa_rev}.
\\
Additionally, there was a typo: a coefficient of $2/3$ should be added in Eq.~25 in the original paper like so 
\begin{align}
\frac{C'}{D'}&=\frac{(\overline{a_3})_{\rm L}+(\overline{a_3})_{\rm H}}{(\overline{a_0})_{\rm L}+(\overline{a_0})_{\rm H}}\nonumber \\
&= \frac{2}{3}\left(-0.295\cos\phi_2\sin\phi_2 +0.050\sin^2\phi_2 \right).
\tag{3}
\label{eq:Igamma3_rev}
\end{align}
This typo does not affect the results of the original paper because $a_3$, the curved lines in Fig. 21 in the original paper, was calculated using the correct expression found in Eq.~\ref{eq:Igamma3_rev} in the erratum.\\\\

\bibliography{ngamma}

\end{document}